\documentclass[12pt,preprint]{aastex}
\usepackage{graphicx} %DB: included package
\usepackage{graphicx}
\usepackage{txfonts}
\newcommand{\minusone}{$^{-1}$}

\newcommand{\CV}{\ion{C}{5}}
\newcommand{\CVI}{\ion{C}{6}}
\newcommand{\OVI}{\ion{O}{6}}
\newcommand{\OVII}{\ion{O}{7}}
\newcommand{\OVIII}{\ion{O}{8}}
\newcommand{\NVII}{\ion{N}{7}}
\newcommand{\NVI}{\ion{N}{6}}
\newcommand{\NV}{\ion{N}{5}}
\newcommand{\NeVIII}{\ion{Ne}{8}}
\newcommand{\NeIX}{\ion{Ne}{9}}
\newcommand{\NeX}{\ion{Ne}{10}}
\newcommand{\skipthis}[1]{}

\newcommand{\ps}{{\rm sec}^{-1}}

\newcommand{\kmps}{{\rm km}\,{\rm sec}^{-1}}
\newcommand{\erg}{{\rm ergs}}
\newcommand{\flux}{{\rm ergs\ cm^{-2}\ sec^{-1}}}
\newcommand{\lum}{{\rm ergs\ sec^{-1}}}

\def\micron{\hbox{$\mu$m}}
\newcommand{\be}{\begin{equation}}
\newcommand{\ee}{\end{equation}}
\newcommand{\e}{et al.\ }

\newcommand{\asec}{$^{\prime\prime}$}
\newcommand{\amin}{$^{\prime}$}

\slugcomment{December 31, 2008 -- {\it Accepted by ApJ}}
\shorttitle{Chandra Observations of 73P/B.}
\shortauthors{Wolk et al.}

\begin{document}

\title{Chandra's Close Encounter with the Disintegrating Comets
73P/2006 (Schwassmann--Wachmann--3) Fragment B and
C/1999 S4 (LINEAR)}


\author{S.~J. Wolk}
\affil{Harvard--Smithsonian Center for Astrophysics, 60 Garden
Street, Cambridge, MA 02138}

\author{C.~M. Lisse}
\affil{Planetary Exploration Group, Space Department, Johns
Hopkins University Applied Physics Laboratory, 11100 Johns Hopkins
Rd, Laurel, MD 20723}

\author{D. Bodewits}
\affil{KVI atomic physics, University of Groningen, Zernikelaan
25, NL-9747 AA Groningen, The Netherlands\\Present address: NASA Postdoctoral Fellow, Goddard Space Flight Center, Solar System Exploration Division, Mailstop 690.3, Greenbelt, MD 20771}
%DB 22/10/07: added affiliation.

\author{D.~J. Christian}
\affil{Eureka Scientific Inc., 2452 Delmar St., Oakland, CA 94602}

\author{K. Dennerl}
\affil{Max-Planck-Institut f\"{u}r extraterrestrische Physik,
Giessenbachstra$\beta$e, 85748 Garching, Germany}



%\altaffiltext{1}{Harvard Smithsonian Center for Astrophysics,
%       60 Garden Street, Cambridge, MA 02138}
%\altaffiltext{2}{European Southern Observatory, Karl-Schwarzschild Strasse 2,
%      D-85748 Garching bei M\"unchen, Germany}


\begin{abstract}
On May 23, 2006 we used the ACIS-S instrument on the Chandra X-ray Observatory (CXO) to study the X-ray emission from the B fragment of comet 73P/2006 (Schwassmann-Wachmann 3) (73P/B). 
We obtained a total of 20 ks of CXO observation time of
Fragment B, and also investigated contemporaneous ACE and SOHO solar wind physical data. The CXO data allow us to spatially resolve the detailed structure of the interaction zone between the solar wind and the fragment's coma at a resolution of $\sim$ 1,000 km, and to observe the X-ray emission due to multiple comet--like bodies. We detect a change in the spectral signature with the ratio of the \CV/\OVII\ line increasing with increasing collisional opacity as predicted by Bodewits \e (2007).  The line fluxes arise from a combination of solar wind speed, the species that populate the wind and the gas density of the comet. 
We are able to understand some of the observed X-ray morphology in terms of non-gravitational forces that act upon an actively outgassing comet's debris field.
We have used the results of the $Chandra$ observations on the highly fragmented 73P/B debris field to re-analyze and interpret the mysterious emission seen from comet C/1999 S4 (LINEAR) on August 1st, 2000, after the comet had completely disrupted.  We find the physical situations to be similar in both cases, with extended X-ray emission due to multiple, small outgassing bodies in the field of view. Nevertheless, the two comets interacted with completely different solar winds, resulting in distinctly different spectra.

\end{abstract}


\section{Introduction}

ROSAT's discovery that comets are bright in X-rays has opened up a new window on Solar System physics (Lisse \e 1996, Lisse \e 2001, Dennerl  \e 1997, 2002, McCammon \e 2002, Pepino \e 2004, Bhardwaj \e 2007).   By now, over 20 comets have been studied in X-rays and the physical emission process has been well identified (Cravens 1997, Krasnopolsky 1997, Cravens \e 2002). Highly energetic solar wind ions (e.g., O$^{7+}$, O$^{8+}$, C$^{5+}$, C$^{6+}$) collide with neutral particles in the comet's coma, resulting in charge exchange.  E.g., a solar wind O$^{8+}$ ion captures one electron and results in \OVIII\ emission. An O$^{7+}$ ion that captures one electron and leads to \OVII\ emission.\footnote {In this paper, we are concerned with the output products of the interactions. Thus, ions with superscript notation are in the solar wind, and ions that are emitting are notated in linear notation.}
The solar wind ions are left in an electronically excited state from which they decay through the emission of X-ray line photons (Bodewits \e 2007).  As solar wind ions gradually lose their charge by subsequent electron capture reactions, we expect charge exchange to occur in the Sunward direction relative to the comet with a strength related to the abundance of the neutral gas species available and the flux of the incident solar wind ions.

The details of the emission depend on the properties of both the solar wind (composition, charge state, density and speed) and comet's mass loss rate (Kharchenko \& Dalgarno 2000, 2001, Beiersdorfer \e 2003, Bodewits \e 2004, 2007). Much effort has been expended in world wide laboratories to measure these charge exchange cross sections and to develop cometary X-ray spectroscopy as a powerful tool to monitor the composition, charge state, density and speed of the solar wind as well as to probe comet activity (e.g. Beiersdorfer \e 2003, Bodewits \e 2004), but this work is still ongoing.   In order to understand the solar wind--coma interaction, state selective cross sections and branching
ratios are required to accurately model the de-excitation cascades of the different ions (Bodewits \e 2007).  This is challenging as the total cross sections and the population of specific excited states depend strongly on the collision energy and the species involved.

Periodic comets are small (0.5--10 km) bodies composed of a mixture of non-volatile dust grains and frozen gasses, which traverse the Solar System on elliptical orbits. As they approach the Sun , comets develop an atmosphere (coma) of sublimating neutral gases, mainly H$_2$O, CO (5-20\% vs. H$_2$O) and CO$_2$ (1-10\% vs. H$_2$O) with trace amounts of other molecules. As these gasses expand away from the comet nucleus, solar UV photons  and solar wind ions photo-dissociate and/or -ionize them on time scales of 10$^4$ -- 10$^7$ seconds (length scales of 10$^4$ -- 10$^7$ km) depending on the species. As the comet retreats from the Sun and the incident solar flux lessons in intensity, these time and length scales lengthen by the square of the heliocentric distance.

% Within the bow shock, the cometopause identifies the   transition between the collisionless, but mass loaded solar wind, and the %collisionally dominated, neutral material outflowing from the comet.  In between the bow shock and the cometopause, in the region %known as the cometosheath, a significant amount of momentum is transferred from the outflowing cometary neutrals to the solar wind %ions and the bulk of the mass loading occurs. It is in the cometosheath that charge transfer from the cometary neutrals to the highly i%onized solar wind should be occurring. %This charge transfer is commonly referred to as charge exchange (CXE).

The solar wind interaction with comets is dominated by mass loading of the solar wind by heavy, slow cometary pickup ions (see recent review by Ip 2004).  The cometary ions are produced by photo-ionization and by charge exchange %ionization of neutrals in the cometary coma 
(Cravens 1991). The interaction begins outside the bow shock, which is the primary working surface of the interaction where the %coma 
mass loading becomes large enough to change the macroscopic flow patterns of the solar wind.
%Once ionized, Coulomb repulsion greatly reduces the charge exchange cross-section with solar wind ions. 
Also, such cometary ions are quickly removed by the solar wind.  Models show that the interaction of the expanding coma with the solar wind results in the formation of a kidney-shaped region bounded by a contact surface and inner shock, where the dynamic pressure of the expanding cometary gasses balances the dynamic pressure of the solar wind (Kabin \e 2000). Similar to the termination shock of the heliopause, this is the zone where the solar wind becomes highly mass loaded by picking up ionized local species
(Gringauz \e 1986). 

The bow shock forms the outer boundary of this interaction region. Before X-ray observations became available, this complex interaction zone, was poorly studied only along single chord trajectories through the coma by {\it in situ} energetic particle analyzers in three comets: 1P/1986 (Halley), (Giotto -- Krankowsky \e 1986; Vega 1 and 2 -- Galeev \e 1986) 26P/Grigg-Skjellerup by Giotto (Johnstone \e 1993) and 19P/Borrelly by DS-1 (Young \e 2004).  On the other hand, X--ray emission directly probes the entire interaction zone and hence both the macroscopic and microscopic  processes that play a role in shaping of the coma and ion tails, as was demonstrated by Wegmann \& Dennerl  (2005). 

\begin{figure}
\plotone{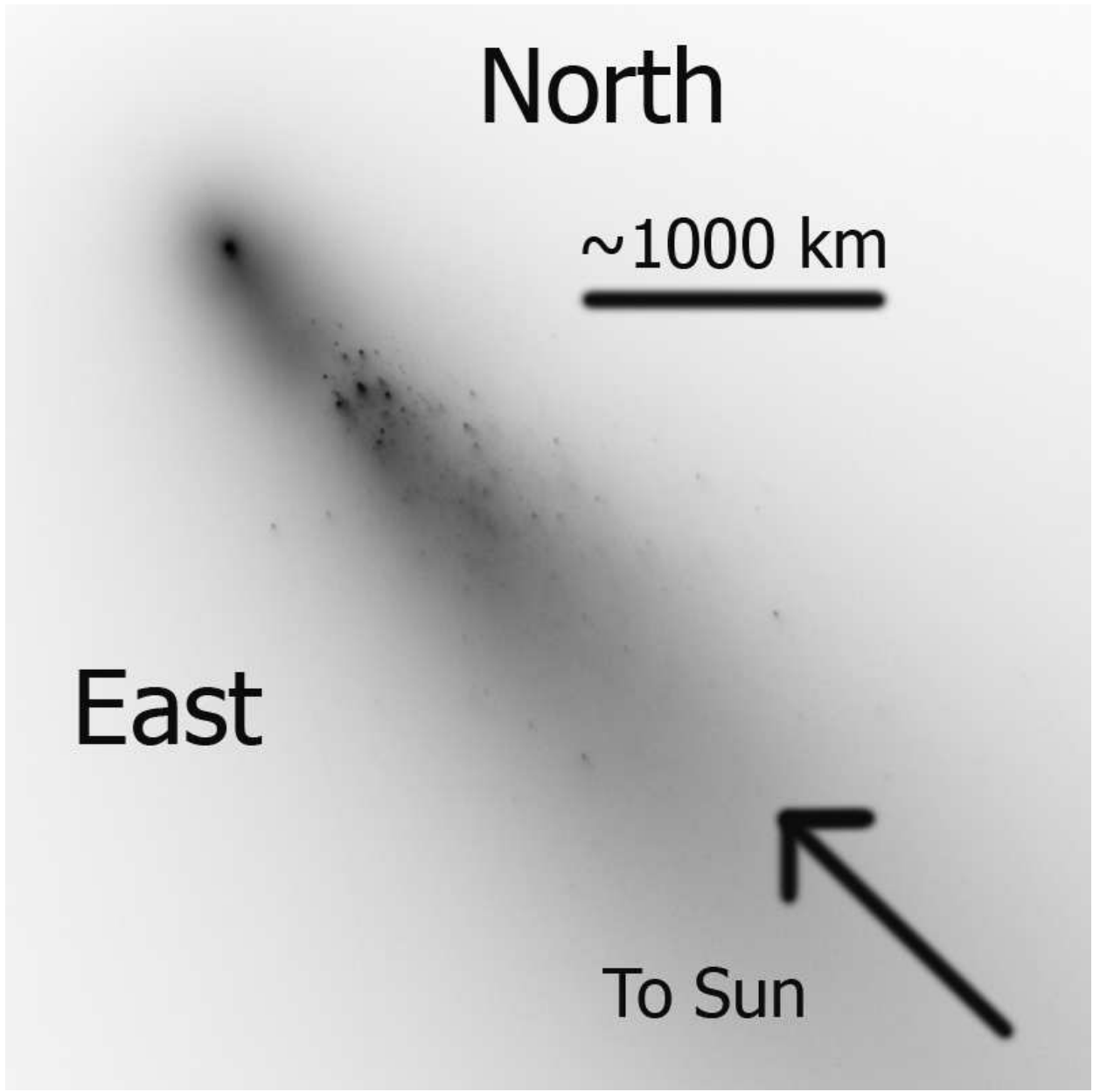}
\caption{HST image of 73P/Schwassmann-Wachmann 3 B from 18 April 2006 using F606W on the ACS/WFC.
North is up and east is to the left.  Multiple fragments trail the
main fragment with the bulk to the south of the Comet--Sun line, which
was nearly aligned with the cometery motion. 
Image by Weaver \e
(2006; http://hubblesite.org/newscenter/archive/releases/2006/18/image/a/.) \label{HST}}
\end{figure}

Comet 73P/Schwassmann-Wachmann 3  was a small short period comet that broke up into several pieces in 1995 (Boehnhardt \e 1995). Three main pieces (B, C, and E) survived this breakup to re-appear during the 2006 apparition. Each of these pieces developed its own coma.  %As a result of this split,
The fragments further split into at least 43 major sub-components (as tracked by JPL -- cf. $Horizons$\footnote{http://ssd.jpl.nasa.gov/horizons.cgi}) and literally thousands of other bodies too small to be detected, except during the initial flaring produced during their creation and ejection. Figure~1 shows a Hubble image of Fragment~B. The fragments were found to have similar chemical compositions in their emitted gas and dust (Schleicher 2006, Villanueva \e 2006, Dello Russo \e 2007, Sitko \e $in~prep.$) but different mass loss rates, depending on their detailed structure. In terms of activity and outgassing, this comet was weak, comparable to Encke in November 2003 (Q$_{gas} \sim 10^{28}$ mol.~$\ps$; Lisse \e 2005, Wegmann \e 2004). On the other hand, 73P/B had an extremely close approach to Earth in May 2006 (0.07 AU), about a factor 5 closer than Encke in 2003.  Due to its high activity at the time -- it was in outburst during May -- Fragment~B was one of the two brightest fragments of the comet 73P/Schwassmann-Wachmann 3 debris train. 
The HST image  (Figure~\ref{HST}) shows structure analogous to comet
C/1999 S4 (LINEAR; hereafter LS4) as seen after its break-up in July - August 2000 (Weaver \e 2001). The morphology of both comets shows debris spread in a triangular structure spreading along the comet-Sun  line.

Due to its proximity and break-up, the world--wide study of the comet
73P/Schwassmann-Wachmann-3 was extensive. $Chandra's$ contribution, discussed in this paper, is unique owing to its well calibrated sensitivity to photons with energy  above 300~eV, superior spatial resolution and spectroscopic capability. Several other comets have now been studied with Chandra with which to compare our results.
In the case of SW3, $Chandra$ observed the most variable large fragment,
Fragment~B (73P/B) -- from which over 35 other tracked fragments have been attributed, % ref http://www.oaa.gr.jp/~oaacs/nk/nk1303.htm
while XMM-Newton, SWIFT and Suzaku observed Fragment~C, most likely the parent nucleus of the active fragmentation. %It is unclear whether any fragment of the comet will complete another orbit around the Sun.

This paper discusses the $Chandra$ observations of 73P/B.  There were several crucial aspects which motivated this observation.  First,  was the very close approach of the comet to the Earth.  The nine comets observed with Chandra have ranged in spacecraft-comet distance  from 0.26 AU (8P/Tuttle; January 2008) to over 1.6 AU (Holmes; October 2007) with a mean distance of  0.71 AU.  At a distance of 0.1 AU, 73P/B held the possibility of unprecedented spatial resolution as well as a minimum of difference between the state of the solar wind as measured by near-Earth satellites and that experienced by the comet.  Second, was the expected low outgassing rate of the comet.  This allowed for the possibility of identifying changes in the spectral signature across the coma.  Spectral changes as a function of position were predicted by Bodwits \e 2007, but have not been seen.  Third, we hoped to better understand the disintegration process the comet was undergoing by investigating the interaction of the neutrals emanating from the comet with the solar wind.

In the next section, we discuss the basics of the observations, including the solar wind data, the X-ray data, the basic X-ray morphology and temporal variations.  In \S 3, we discuss the spectral fitting using a CXE constrained model.  We find that the CXE fit works,  but that there is an anomaly at low energy introduced by lines which are below our fitting energy. We discuss the implications of our observations and model fits  discussion in \S 4.  We start by discussing, the overall morphology -- including density enhancements and  the gas production. Starting in \S4.3 we delve into the details revealed by the spectroscopy, including the state of the solar wind as ascertained from the X-ray data and the spectral structure of the comet. We close the discussion in \S 4.5  by comparing 73P/B with other comets observed in the X-ray, first by comparison of the X-ray to bolometric luminosities among several comets, and then by compare the X-ray observations of 73P/B to the previously unanalyzed post break up observations of Linear/S4. In \S 5, we summarize our conclusions.

\section{Observations}
\subsection{X-ray}
$Chandra$ observations of comet 73P/B were obtained on 23 May 2006 in 6 hours (20 ksec), using the ACIS-S CCD spectral array, which combines the abilities to image X--rays at a platescale of 0.5\asec/pixel, and to produce well calibrated, moderate resolution spectra ($\Delta$E $\sim$ 110 eV FWHM, $\Delta$E$_ {Gaussian}$= 49 eV) from 0.30 to 2.0 keV. ACIS has 6 active CCDs.  % I2,I3, S1, S2, S3 and S4. 
Each CCD as an 8\amin\ field of view. The gap between the chips was 22\asec\ in the horizontal direction and 163\asec\ in the vertical direction.   Our analysis, and the aimpoint of the telescope, focused on the back illuminated S3 chip which has better low energy sensitivity than the other chips.
 During the $Chandra$ observation 73P/B was 0.97 AU from the Sun and 0.10 AU from the
Earth, with a total visual magnitude of V $\sim$ 7.5 (Figure~\ref{lightcurve}),
%ref http://www.aerith.net/comet/catalog/0073P/2006.html
a total luminosity  L$_{optical} \sim  7.25 \times 10^{17}$ erg sec$^{-1}$, and an estimated Q$_{gas}$  of $2 \times 10^{28}$ mol.~sec$^{-1}$ (Schleicher, 2006)  moderately active for an ecliptic comet.

We obtained 5 pointed CXO observations of 73P/B on 23 May 2006 from 05:30 to 11:45 UT each of $\sim$~65 minutes (3.7 ksec).  Data were taken in ``very-faint" (VF) mode with a 3.2 second frame time and an energy filter which excluded energies above 13~keV to prevent telemetry saturation.  %On downlink data with grade values of 1, 5 and 7 were excluded, as they are generally not X--rays. 
Standard processing applies a very faint filter to exclude additional events that are generally not caused by XÐrays. On downlink
data with grade values of 1, 5 and 7 were excluded, as they are generally not XÐrays.
%standard processing applies a very faint filter to exclude additional events that are generally not caused by X--rays.

The method of $Chandra$ observation was kept simple: the ACIS-S array was pointed on the sky 2\amin\ in front of the comet along the direction of its apparent motion, and the comet was allowed to move through the field of view.  No active guiding on the comet was attempted. The predicted location of the B nucleus crossed the aimpoint 2\amin\ from the northwestern edge of the S3 chip 2 ksec seconds into each observation. Thus, the nucleus was only on the S3 CCD about 75\% of the exposure.  After 2~ksec seconds the telescope was repointed by about 8 arcminutes to the south-southwest and the process repeated.  Repointing took approximately 400 seconds.  The direction to the Sun is 21 degrees south of due west.  Fragment C, which indicates the orbital direction of motion, is in the direction 71$^{\rm o}$ south of west  (dRA*cosDec/dt = 357.0\asec/hr.,
d(Dec)/dt = -222.1\asec/hr.). 
% D physcal diamter = 2!pi distance angular size /360
%D = 2!pi 15,000,000 (km)  0.0666667/360  % 4'
%4' =17,500km == 12,000 at 0.07 AU

$Chandra$ observations return a list of time-tagged detections of individual photon pulse heights and spatial locations. %CXO was able to follow the comet's %non-sidereal motion using multiple pointings, and 
%The target was centered in the back-illuminated CCD chip S3, the most sensitive CCD for energies below 1 keV.
%No obvious bright background sources were detected in our data, other than the comet. %(Figure~\ref{full}). 
After removing three noted point sources, we constructed a comet image with the photons re-mapped into a coordinate system moving with the comet's nucleus using the ``sso\_freeze'' algorithm, part of the $Chandra$ Interactive Analysis of Observations (CIAO) software.  The layout of the observations on the sky and reprojected into comet-centric  coordinates is shown in Figure~\ref{LAYOUT}.
The resulting effective field of view of the back illuminated S3 chip in the comet-centric coordinate system was about 10\amin\ $\times$ 10\amin.

%We also constructed an exposure map corrected version of this image using new capabilities available in CIAO~4.

A total of 1865 photons were detected from the comet in the 300 to 1200 eV range in the central elliptical emission region. %(Figure~\ref{full}).
The average count rate in the 300 to 1200 eV energy range over the 20.8 ksec of observations was 0.13 cps, similar to the 0.1 cps rate found for comet Encke in 2003 (Lisse \e 2005).
Spectra were compared to a solar wind velocity sensitive Charge-Exchange (CXE) model (Bodewits \e 2007) using the XSPEC fitting routines (Arnaud 1996).  This model is discussed in more detail in \S3.

\subsection{Solar Wind}
Because of its close approach to Earth -- relative to other comets observed at X-ray wavelengths -- comet 73P/B allowed us, for the first time, to link observed X-ray spectra from the coma with the directly measured physical conditions of the solar wind which enter the interaction zone. We obtained solar wind data from the on-line data archives of ACE (proton velocities and densities from the SWEPAM instrument, heavy ion fluxes from the SWICS and SWIMS instruments\footnote{http://www.srl.caltech.edu/ACE/ASC/level2/index.html}) and SOHO (proton fluxes from the Proton Monitor
Instrument\footnote{http://umtof.umd.edu/pm/crn/}).
Both ACE and SOHO are located near Earth, at its Lagrangian point L1, at 0.01 AU from the Earth.. In order to map the solar wind from L1 to the comet position, we used the time shift procedure described by Neugebauer \e  (2000). The calculations are based on the comet ephemeris, the location of L1, and the measured solar wind speed. With this procedure, the time delay between an element of the co-rotating solar wind arriving at L1 and the observed X-ray emission can be estimated. During our observations, solar wind observed at L1 arrived at the comet within less than 5 hours.  For comets previously observed by $Chandra$ the typical  delay was over 41 hours. %-- ranging from 9.1 hours for Tempel 1 to over 6.6 days for C/1999 T1 (McNaught-Hartley). 
The solar wind proton velocity and heavy ion composition around the observations are shown in Figure~\ref{fig:sw}.

%PARAGRAPH ON SPATIAL RESOLUTION
\subsection{Morphology}
At a distance of 0.1 AU the resolution of $Chandra$ equates to a spatial resolution of about 50 km per pixel. Since our X--ray observations were photon starved, smoothing is required to improve S/N. Overall there were about 1000 photons between 350~eV and 1.2 keV detected within 100\asec\ of the nominal location of the comet nucleus. This is an area of about 30,000\asec$^2$ so only 1 photon was detected per box 5.5\asec on a side.  The background is about 0.14 count per second on the S3 chip.\footnote{$Chandra$ Proposers Observatory Guide (POG) Version 9.0 \S6.15 Table 6.10} To obtain 3 $\sigma$ noise statistics, we require at least 10 counts/pixel.  This provides an effective pixel of about 18\asec\ $\times$ 18\asec\ (or
a spatial resolution of about 1,000 km).

The ACIS detector has two back-illuminated (S1 and S3) and four front illuminated (S2, S4, I0, I1) CCDs.  The advantage of the back--illuminated CCDs is that they have greater sensitivity below 800 eV and thus are more sensitive to the dominant CXE lines at the expense of somewhat lower energy resolution. 
Chips S1 and S3 show the strongest apparent emission primarily due to their high sensitivity below 500 eV (Figure 3).   However, the S1 background is known to be higher than S3 and
These values are indicative of a somewhat larger excess on S3 because the background level of S1 is about 25\% higher than S3 and we expect the background to be about 1000 counts higher on S1.\footnote{Ibid.}
 %Overall, there is no evidence that the cometary charge exchange emission extends beyond the boundaries of S3.

 %The S3 CCD detected less than 100 photons more than S1 in the 300~eV to 1000 eV passband (Table~1).  These values are indicative of a somewhat larger excess on S3 since the background level of S1 is about 25\% higher than S3 and we expect the background to be about 1000 counts higher on S1.\footnote{Ibid.}
%In general, the front illuminated CCDs, excluding S4\footnote{S4 is known to have an higher background that the other front-illuminated  CCDs -- 0.16 cps vs. 0.10 cps average for I0-I3 in the 0.5-2.0~keV range.- Ibid.}, should have about the same background.
%Overall, there is no evidence that the cometary charge exchange extends beyond the boundaries of S3 in the direction orthogonal to the Sun, but not in the anti-Sun direction.  However, spectroscopic analysis beyond the scope of this paper would be needed to confirm this.

Concentrating on the morphology of the data on the S3 chip, we restrict the data to nominal energies below 1000 eV and above 300~eV, where the bulk of the detectable CXE lines have been found for other comets. 
To create a visualization which reflects the 18\asec\ resolution, the data were smoothed to create an image using the CIAO tool {\it csmooth} to a resolution of about 12\asec\  $\times$ 12\asec. A second version of the {\it csmooth} algorithm was run at a resolution of about 8\asec $\times$ 8\asec.   A composite of these images is shown in Figure~\ref{fig:image}. The smoothed images show the primary emitting region to be about 25,000 km in extent in the Sunward direction and about 40,000 km in the orthogonal direction.
The peak of the emission is not significantly displaced from the nominal B nuclear location.
There is a second peak about 90\% the strength of the nucleus about 50\asec\ to the southwest, or about 2.5 resolution elements off of the direction of motion, in the forward direction. The 73P/B emission is almost entirely contained on the S3 chip - there is no evidence of CXE signal on the S2 and S4 chips (Table 1). 

Overall we identify five structures:
1) the background -- essentially the corners of S3;
2) an outer ellipse with its minor axis oriented along the direction of motion;
3) a triangular polygon within the emitting region;
4) the nominal nuclear peak (which is {\em not} detected as an enhancement); and
5) a lesser peak displaced from the nominal nuclear peak by about 50\asec\ in the direction of fragment 'C'.
To quantify the various structures we calculated the counts in each region.
The general emission is widely dispersed over about 50,000 km roughly east to west and roughly 75,000 km north to south. The brighter emission (triangle) is displaced north of the main emission but still south of the nucleus.
In Table~\ref{SurfBright} we calculate the surface brightness in counts per pixel of each region and the brightness of each region above the overlaying region.  From this table, it is clear that the oval region is a true region of excess emission above the background and that the triangle is clearly brighter than the ellipse. There is 98\% confidence that the nuclear peak is real and about 85\% confidence in peak 2.  It is important to note that this sort of complicated X--ray structure has been observed for only one other comet to dateâ Linear/S4 (Lisse \e 2001).

In the broadest energy range, 300-1200~eV, 73P/B appears roughly elliptical in shape, with the major axis aligned with the direction of apparent motion (see the left hand side of Figure~\ref{LAYOUT}).
Data in the softest bands, 350-500~eV -- which should be dominated by emission from Carbon and %Nitrogen ions (\CIV, \CV, \NV, \NVI), peak sharply near peak 3, with emission extending to the northern apex of the triangle and an additional emission along the eastern edge of the triangle, reflecting much of the same structures.  
Data in the middle band, 500-700~eV -- which should be dominated by emission from ionic Oxygen -- show three concentrations of emission near the nucleus.  A central one associated with the nuclear peak  and peak 2, one to the south along the southern edge of the triangle and a third one near the northern apex of the triangle.
Harder channels show a strong peak at the southwestern apex of the triangle, with little additional coherent emission. An interesting preliminary result is the relatively limited area of overlap among the emission regions when photons are sorted by energy.  This indicates that the Oxygen emission arises from different regions than the Carbon emission.
%While faint, the little peak to the south is several $\sigma$ in significance.

Since the morphology driven by the dynamical behavior of solid material emitted from a comet's nucleus, we need to consider the modifying effects of solar radiation.  %Once in the tail, the forces acting on cometary dust are charged particle/electromagnetic interactions with the solar wind, solar gravity, and solar radiation pressure. 
We ignore the effects of the solar wind (charged particle energy deposition, spallation, charging, and E-M forces from the fields embedded in the wind), as these are negligible for particles of radii $>$ 0.1 \micron\ (Burns \e 1979, Burns 1987, Horanyi and Mendis 1986, 1991). 
Solar gravity treats the emitted dust the same way it treats the comet's nucleus, i.e. the dust tracks the comet's Keplerian orbit (with typical heliocentric velocities ~10 - 50 km/sec), modulo the minor perturbation from the terminal velocity (0.001 - 0.5 km~$\ps$) imparted by the emission process and expanding coma gases (Lisse \e 1998, 2004).  Solar radiation pressure shows its effect as acceleration in the anti-solar direction (an outgassed  particle effectively sees a lower mass sun than does the nucleus), affecting all particles, proportional to the product of the surface area multiplied by the photon absorption-scattering efficiency/mass. The most pronounced effect is on small particles with size $\sim$1.0 \micron\ with high surface area/mass ratios and high resonant scattering efficiency for the optical photons that dominate the Sun's spectrum by number.

For small particles, the parameter $\beta \sim 0.47/(\rho \times a$[\micron]) is used to evaluate whether the effect of gravity or wind pressure will dominate the trajectory of a particle.
The limiting cases help us bound the range of dust particle dynamical behavior. 
When $\beta \sim$ 1, the  
forces on dust are large enough that it moves down and out of the comet's tail within hours to days.
For the $\beta > 1$ ($\sim$ 0.1 - 1 \micron\ radius) particles, the net acceleration is so large that the anti-solar direction is approximately constant over the few days that the dust is present in the tail, and the dust trajectory is roughly linear in the anti-solar direction (see the anti-solar structure evident in the HST images of 73P/B, Figure 1, and in the multiple tail structures of the Spitzer 24~\micron\ image of the fragment train of Reach \e\footnote{http://www.spitzer.caltech.edu/Media/releases/ssc2006-13/index.shtml}). For $\beta < $0.001 (radius $>$ 500~\micron), the net acceleration in the comet frame is very small, the particle remains in the vicinity of the nucleus for many months, and the particles slowly spread out along the comet's velocity direction due to the Keplerian shear imparted by the velocity of expulsion (see the line of emission from large dust lying along the comet's orbital path and `connecting' the individual fragments in the Spitzer 24~\micron\ mosaic of Reach \e\footnote {Ibid.}).  The abundant intermediate sized particles with 1 $ >  \beta >$ 0.001  (1 - 500~\micron\ radius), which contain much of the emitted mass from 73P/B, follow trajectories between these two extremes (c.f. Fig.~6 of Reach \e 2000 and Reach \e 2007). We thus expect to find them, and any gas emitted by them, to lie in a spatial region between the anti-solar direction vector and trailing direction of the comet's motion. 
This interpretation is complicated by the charge exchange mechanism that emits a photon for each interaction with a released neutral.  Once released the neutrals travel in random directions and have been seen to persist over 100,000 km away from their source (c.f. Tempel~1 - Lisse \e 2007). Hence the emission should, and does, fill the field.  Any signature of multiple emitting sources would have to be seen above this overall signal.
While the signal to noise in Figure~\ref{fig:image} cannot support this interpretation, the bulk of the  flux is seen either coincident with the nucleus or in the direction towards the Sun, relative the nominal nuclear position.

\subsection{Temporal Variation}

Temporal variation has been previously studied in several comets. Impulsive events were seen in several X-ray observations of comets (Lisse \e 1996, 1999, 2001; Hyakutake, Encke and Linear S4 respectively). X-ray observations of the 2003 passage of Encke ( Lisse \e 2005) indicated a $\sim$11.1 hour periodicity attributed to nuclear rotation.  Observations of Tempel 1 associated with Deep Impact (Lisse \e 2007) showed not only the flare associated with the impact but a quasi-periodic brightening with a two-week period.  For 73P/B temporal variability could also provide 
another line of evidence about the reality of the features. We examined four time slices of the data, each covering about 5000 seconds. Conclusions from these time slices include:
\begin{enumerate}
\item The nuclear region is always bright.
\item Peak 2 is always brighter than the surrounding material.
%\item Peak 3 was transient with flux only during the middle 10 ks.
\end{enumerate}
%Hence Peak 3 may not be associated with a permanent (on timescales of hours), distinct physical object. 

\skipthis{
Overall, there were 6679 photons between 300 eV and 1.2 keV detected on the S3 array, 1865 within the central ellipse region ``Ellipse~0'' which covered about a quarter of the region of space observed by the S3 chip.
The motion of the comet across the edge of the chip complicated timing analysis. 
The whole of Ellipse~0 is only on the entire CCD briefly.
The somewhat more intense and compact triangle of emission has a longest extent of about 210\asec.  It is completely on S3 for the final $\sim$ 70\% of each observation.

 Correcting for the periods between pointings and time while portions of the triangle were off the S3 CCD, the total observing time was  16,400 seconds.  A total of 740 photons were detected in the region between 300~eV and 1.2 keV.
This yields a mean count rate of 0.045 $\pm$0.002 counts per second.
Comparing the 5 separate data sets we find the count rate varied from 0.040 to 0.053 ($\pm$0.004). The plotted light curve shown in
Figure~\ref{fig:lc} shows limited variability, peaking during the middle observation with the smallest count rate during the last observationThe ACE and SOHO data indicate that the proton density was fairly steady during the four day period surrounding the day 143.25-143.5 observation. The solar wind velocity was fairly steady ($\pm25$ km s\minusone) during the 20 hour observation. 
}

We studied flux changes to look for variations on shorter time scales.  The mean count rate  in the central ``triangle" region was about 9.0 $\pm$ 3.0 counts per 200 seconds.  Of the samples unaffected by repointing, one was high by (almost exactly) 3$\sigma$ from the mean flux rate.
Another sample was remarkably low by 2.7$\sigma$.  Neither deviation was adjacent to a second deviation of similar magnitude.
Hence, there is no evidence for temporal fluctuations of the X--ray flux larger than 10\% during the 6-hour period of observations.

\section{Spectral Fitting and Results \label{sec:spectra}}

We extracted spectra from several regions centered on the comet with a minor axis along the comet-Sun line in the S3 chip, and from several apertures on the adjacent S2 and
S4 chips. %, to investigate if they had any emission. 
The apertures for the S3 chip included: (1) a large box shaped aperture to include as many counts as possible for a total, high signal-to-noise spectrum (bigbox), (2) approximately annular ellipse at increasing larger radii from the nucleus to find any changes in the spectrum (these are called Ellipse~0, Ellipse~1 and Ellipse~2 - see Figure~\ref{fig:image}) and (3) the triangular region mentioned in \S2.3 (Morphology).  We do not individually examine the point-like peaks due to their low intrinsic brightness. 

The spectra were extracted using the standard
CIAO (v3.4) tools, including the creation of the Ancillary Response Function and Redistribution Matrix Function files that compensate for local and temporal variations in the ACIS response. While the regions are extracted in the comet's frame of reference, the analysis is done in fixed chip coordinates so that the appropriate responses are used.  Background spectra were extracted from comet emission-free regions from the anti-Sun side of the S3 chip.

%Two types of spectral models were fitted: 1) an emission line model used previously for our analysis of comet Encke (Lisse \e 2005) and Tempel 1 (Lisse \e 2007), and 2) a charge exchange (CXE) model as presented in Bodewits \e (2007). The emission line model has 12 emission lines with the line energies fixed as given in Bodewits et al. (2007) and line intensities as free parameters.  We include the model primarily for The second model is a 

The data were fitted to the CXE model as presented in Bodewits \e (2007).
In this CXE model,  each group of ions in a species were fixed according to their velocity dependent emission cross sections to the ion with the highest cross section in that group. Thus, the free parameters were the relative strengths of H- and He-like Carbon (299 and 367 eV), Nitrogen (\NV\ at 419 and \NVI\ at 500 eV) and Oxygen (\OVII\ at 561 and \OVIII\ at 653 eV). Additional Ne lines, the \NeIX\ forbidden, inter-combination and resonance lines around 907 eV and the \NeX\ Ly-alpha at 1023 eV were also included  in our CXE model, giving a total of 8 free parameters. Spectral parameters were derived using the least squares fitting technique with the FTOOLS XSPEC package (Arnaud \e 1996).

In the model, the lines were assumed to be intrinsically sharp, only broadened by the instrumental resolution ($\sim 50$~eV).
The spectra were fit in the 300 to 1000 eV range. This provided 49 spectral channels, and thus 41 degrees of freedom. Difficulties in modeling ACIS spectra below 300 eV have already been discussed in Lisse \e (2001, 2007), as a result of the rising background contributions and decreased effective area near the instrument's Carbon edge. Thus, fitting below 300~eV was not be attempted here.

A sample spectrum fit with the full CXE model is shown in Figure~\ref{fig:3specstuff}. We show the spectrum of the largest aperture centered on the S3 chip and the residuals to the fit and input CXE model.
This spectrum showed strong emission lines of \CV\ (299 eV), \CVI\ (367 eV), \NVI\ (419 eV) and \OVII\ (561 eV) as seen in previous comets (Lisse \e 2001, 2005, 2007, Krasnopolsky \e 2002, Bodewits \e 2007). The \NVII\ at 500 eV was not significant and there were 1 to 3 $\sigma$ detections of \NeVIII\ (907 eV) and \NeIX\ (1023 eV) lines.
Spectral fitting results for the full CXE model with a solar wind velocity $v  = 450$ km sec$^{-1}$ are summarized in Table~2. % --\ref{tab:spec_point}.
Fluxes are given in units of ergs cm$^{-2}$ sec$^{-1}$.
Errors are obtained by averaging over the calculated $\pm$90\% confidence contours (corresponding to $\chi^2$ =2.7 or 1.6 $\sigma$).
Goodness of fit parameters, fluxes and the most significant line ratios are also listed. The overall X--ray flux from the comet was      $8.0\times 10^{-13} \flux$, corresponding to  a total X--ray  luminosity from 0.3 - 1.0 keV of $2.3\times 10^{13} \lum$.  About 25\% of the flux comes from a small triangular region located forward of the nucleus relative to the direction of motion. 

The columns of Table~2 are arranged spatially. The first fit is for a small circle 100 pixels in diameter centered on the nominal nucleus. The next three fits are cylindrical shells of elliptical cross section moving progressively outward. The final column lists fitting results to the  C/1999 S4 (LINEAR, hereafter LS4) post-breakup spectra (extracted for the same large square ``bigbox'' region as SW3), which we will compare to our SW3 results in \S 5.

We have searched for changes in the spectra as a function of distance from the nucleus.  In Figure~\ref{fig:specpoly}, we compare spectra from the inner region and the very central 100 pixels centered on the nominal nucleus.  The spectra are scaled by the area of each region and show more relative total counts in the inner region. The innermost spectrum shows very weak emission with \OVII\ being the most significant. Spectra from further away from the nucleus show an increased amount of Oxygen emission, with the \OVII\ emission increasing a factor of 4 to 5 and the \OVIII\ emission doubling from the middle to outer region.

Figure~\ref{fig:specpanda} compares outer, middle and inner ellipse spectra in the 300 to 1000 eV range.  The relative strengthening of the Carbon feature is clearly visible.  We discuss the ion line ratios as a function of distance from the nucleus and CXE cross--sections in \S 4.2.

%END NEW JUNE 1
\section{Interpretation and Discussion}

73P/B is (or perhaps was) a unique object for study in three critical ways. It passed very close to the Earth, presenting us with an unequaled opportunity to study the morphology at high spatial resolution. Secondly, it was a weakly outgassing object, possessing a collisionally thin (to CXE) coma, allowing us to test CXE physics in a collisionally thin medium. Finally, it was fragmenting as we were observing it, allowing additional insight into the evolution of comets. Its X-ray morphology (see Figure 4 and \S 2.3) clearly differs from other comets observed by $Chandra$ in that it lacks either the pronounced umbrella shape of the strongly outgassing comets [Hyakutake (Lisse \e 1996), pre-break up SL4 (Lisse \e 2001), McNaught-Hartley (C/1999 T1; Krasnopolsky \e 2002) or Ikeya-Zhang (153P/2002; Dennerl \e $in~prep.$)] or the ``strong point source with faint diffusive halo of emission'' [Encke (2P/2003; Lisse \e 2005); Neat (C/2001 Q4; Sasseen \e 2006); Tempel~1 (9P/2005); Lisse \e 2007]. Here we find that  the key reason for the differences seen between 73P/B and other comets was the fragmentation process, releasing and diffusing solids that spread out extensively and {\em then} sublimated.

\subsection{Morphology}
The X-ray morphology of 73P/B is complicated by the non-uniform exposure due to the stare and shift observing approach on a moving solar system object. For this work, we have utilized a new capability in CIAO 4 to create an exposure map corrected X-ray image of 73P/B.

The detail of this is shown in Figure~\ref{DETAIL}. While the triangle of emission is still a clear feature in the exposure corrected map, the strongest emission is towards its northern vertex. The bulk of the emission comes from roughly five regions of similar flux located either side of the cometâ Sun line. The situation is further complicated by the on-going disintegration of the comet. This means that non-gravitational forces have a pronounced effect on the position of the B-fragment main mass. This may be the reason why the brightest region is not exactly at the aimpoint, but displaced by about 12\asec.

It is relatively straightforward to understand why the 3 brightness centers lie along the orbital velocity direction. This emission is coming from gas sublimating from large chunks of cometary material, at least 0.1 km effective radius or larger, released at small relative velocities in the recent past. For such bodies, non-gravitational forces such as asymmetric sublimation and radiation pressure are either insignificant or have not had enough time to act appreciably. Similar to the larger chain of $\sim$50 fragments observed (C-B-G-etc.; c.f. Reach \e 2006), these chunks spread out slowly along the orbital direction of motion of the comet due to Keplerian shear (i.e., the small differences, on the order of 1 part in 10$^3$, of their heliocentric velocities).

It is more difficult to understand why the extended diffuse emission appears to be in the Sunward direction. If the extended gas was coming from many icy fragments in a debris field, as suggested by HST observations of the 73P/B tail in April 2006 (Weaver \e 2006), then these should be swept in the anti-solar and anti-orbital velocity direction from the chunks (c.f. Lisse \e1998). The debris seen in the 73P/B HST time lapsed imagery are clearly moving away from the Sun and away from the direction of motion of the SW3 large fragments (see Figure~1). If particles were the source of the CXE emission, the triangle of emission would be rotated about 45 degrees so that one side would be along the comet--Sun line away from the nucleus relative to the Sun. We can also rule out X--ray emission from small icy dust particles, on the order of 0.1 - 10 \micron\ in size, releasing appreciable amounts of neutral water gas as they are accelerated down the cometÕs tail by radiation pressure. %In an extended range down to
%$\sim$  1\micron\  the icy dust particles are being blown outward by solar radiation.
Interestingly, the post disintegration data from comet Linear S4 do show this pattern (see \S4.5).

What is seen instead is emission in the solar direction, following along the chain of debris marked by the 3 brightness centers. One possible explanation for this is being due to overlapping emission from collisionally thin comae centered on each of the chunks, like several small comets strung out along a line. The emission of neutral gas (mainly water) is preferentially towards the Sun, because the Sun warms up this part of the comet the most. % and the solar wind encounters the neutral volatile atoms released from each chunk on its sunward side first.

\subsection {Estimating the Neutral Coma Gas Production}

The X-ray emission of comet 73P/B can be used to make an estimate of the comet's bulk gas production rate.  In the case of 9P/Tempel~1 after the Deep Impact experiment of 04 July 2005, an extension was seen in the $Chandra$ images, identified with the icy material excavated from the comet by the impact  (Lisse \e 2007).
The statistical significance of the extension was $\sim 5\sigma$ above background. Larger scale imaging by the ROSETTA spacecraft over the course of the next few days confirmed the progression of the water gas and OH ejected by the impact (Kueppers et al.\ 2005). The estimated propagation velocity of the extension was 0.5 to 1 km sec\minusone.  We use the estimated total mass of water ejected by Deep Impact ($\sim 5\times10^6$ kg) and the total on-target time of observation for Tempel~1 on July 4-5 of 130 ks to estimate the limiting sensitivity of $Chandra$ to cometary activity as $5\times10^6$ kg/130 ksec (5$\sigma$, equivalent to $\sim 10^{26}$ mol sec$^{-1}$ of H$_2$O at 1.5 AU).
This sets the lower limit on the observed neutral (H$_2$O, CO, CO$_2$ plus minor constituents) outgassing rate from 73P/B for the same solar wind conditions.

  We can estimate a straightforward outgassing rate since the nominal detection of 73P/B was about 25$\sigma$ (Table~1).   The observed outgassing rate is $\sim 3 \times 10^{27}$ molecule sec$^{-1}$ ($\pm$ 50\%).  This is consistent with the 73P/B water production rate values of $1.0\pm 0.2\times 10^{27}$ and  $7.2\times 10^{27}$ measured about 1 month before the $Chandra$ observation (Villanueva \e 2006 and Schleicher 2006  respectively). On the other hand,
  %ApJ 650L90 /IAU CBET 491
it is less than the  $2 \times 10^{28}$ mol.\ sec$^{-1}$ ($\pm$ 30\%) measured by D. Schleicher ($private~communication$) using narrowband optical photometry on 17-18 May 2006. %He measured, a value of  $2 \times 10^{28}$ mol.\ sec$^{-1}$ ($\pm$ 30\%). 
Schleicher also reported the water production rate decreasing by 25\% between the two nights. This  makes $10^{28}$ mol sec$^{-1}$ an upper limit as the activity level on the comet was decreasing monotonically from 18 May to 23 May.  
The field of view used to determine the water production rate of Villanueva, Mumma, et al. is small, on the order of a few arcsec in radius, while that of Schleicher \e is on the order of 30\asec\ - 1\amin. The field of view is important because this is not a comet which produces gas only at the nucleus, but is rather more extended due to the ongoing break--up.  Hence, we expect to have much larger water production estimates from the Schleicher e\ group, and these estimates are on roughly the same spatial scales (a few arcminutes) as determined from analysis of the $Chandra$ images. 
  %Our observations were five nights later and that 73P/B had undergone
  %an outnurst before the  Schleicher 
  %H$_2$O  decrease  (D. Schleicher private communication). But 
  %but this is not
 % surprising.  
% 7/12 letter from M Sitko

There were some differences in the observing conditions between Tempel~1 and 73P/B. The most significant  were the solar wind proton density and ionization state.   The solar wind speeds experienced by Tempel~1 over the course of 30 June --14 July 2005 varied between 400 and 600 $\kmps$, with a median very close to the 450 $\kmps$ value of the solar wind during the relatively brief 73P/B observation.  %Hence, for the same solar wind conditions as Tempel~1 - hile the wind velocities were similar, 
The proton density for the week of the 73P/B observation of about 3 protons cm$^{-3}$ was lower than the mean density of 8 protons cm$^{-3}$ seen the week of the Tempel~1--Deep Impact observation, even though the proton rate was varying by factors of two during the Tempel~1 observation.  
The composition and ionization state of the solar wind was also somewhat different for observations of the two comets. During the observations of 73P/B, the O$^{8+}$/O$^{7+}$ ratio was about half that of the wind during the Tempel~1 observations. As we will demonstrate in the following sections the ionization state of the solar wind can significantly effect the overall X-ray luminosity.  Still the inferred outgassing rate is very modest for a comet. It signifies a low activity
comet, or possibly a very small, active one.

\subsection{Spectroscopy}

\subsubsection{Spectral Classification of Wind State}

%DB 22/10/07 - minor revisions

A survey of cometary spectra obtained with $Chandra$ (Bodewits et al. 2007) suggested a classification based upon three competing emission features, i.e. the combined Carbon and Nitrogen emission (below 500~eV), \OVII\ emission around 565~eV, and \OVIII\ emission at 653~eV.  Firstly, the low energy emission
($<$500~eV) seems to be anti-correlated with the Oxygen emission. This can be explained in terms of the freezing--in temperature of solar wind ions; the charge state distribution of the solar wind reflects the temperature of its coronal source region. Colder source regions hence result in lower charge states. For Carbon, lower freeze--in temperatures imply that more ions are in the He--like state (as opposed to a H--like state).  At these low temperatures,  most Oxygen ions will be in the O$^{6+}$ state, which does not yield any X-ray emission detectable with $Chandra$ via CXE, but UV photons instead. As a net effect, the emission shifts towards the softer part of the X-ray spectrum. This can be seen very clearly in the spectra of 73P/B and 2P/2003 (Encke). In the spectra of those two comets the Carbon/Nitrogen features are roughly as strong as the \OVII\ emission, as opposed to an X-ray spectrum arising from a comet interacting with a hotter heliospheric wind environment, such as the July 14 (day 196) observations of comet LS4 (see Figure~\ref{fig:sw}).  Examining space borne solar wind data confirms that 73P/B indeed interacted with a `cold' wind. Around the time of the $Chandra$ observations, a sequence of three high speed coronal hole streams passed the comet. 
 Coronal holes allow escape from Sun to the solar wind of cold ($\sim 10^6$ K) plasma from low in the corona, near the photosphere, as opposed to the usual outflow of hotter plasma ($\sim 2 \times 10^6$ K) from the top of the corona. 

The ratio between
O$^{7+}$ and O$^{6+}$ ionic abundances has been demonstrated to be a good probe of solar wind states. Zurbuchen et al. (2002) observed that the slow, warm wind associated with streamers typically lies within 0.1 $<$ O$^{7+}$/O$^{6+} <$ 1.0, corresponding to freezing in temperatures of 1.3--2.1 MK. In Figure~\ref{fig:swox}, we show the observed O$^{8+}$ to O$^{7+}$ ratios and the corresponding freezing-in temperatures from the ionizational/recombination equilibrium model by Mazotta \e (1998).  Most X-ray comet observations are within or near to the streamer-associated range of oxygen freezing in temperatures.  The X-ray spectrum from SW/3B is consistent with its being in the warm, slow wind. This contradicts the space borne data indicate a "cold" wind. Hence we believe 
the comet interacted with a corotating interaction region (CIR), a faster stream plowing through the back ground solar wind. It is a mixed state, in contrast to the 'pure' cold polar wind, with which it might have interacted and which would have no O$^{8+}$ at all (Schwadron \& Cravens 2000, Kharchenko \& Dalgarno 2001).

\subsubsection{Solar Wind Abundances and Low Energy Contamination}

The main advantage of 73P/B over previous comets observed in the X-ray regime was that it came so close that (nearly) in-situ measurements of the solar wind were possible, enabling us to perform a direct test of our spectral modeling.
Following the procedure laid out in Bodewits \e (2007), the intensities of the spectral fit can be used to derive spectral abundances by weighting them with the relevant emission cross section. These results can then be compared with ionic abundance data obtained by the ACE/SWICS instrument. %The close proximity of comet 73P/B implied a delay of only 7.2~hours between detection at ACE and interaction with the comet, hence removing much of the uncertainties and disadvantages usually implied by the co-rotational mapping technique (see e.g. Bodewits \e 2007).
 Averaging over the time of the observations, ACE/SWICS measured relative abundances of C$^{5+}$/O$^{7+}$ = 7.6 $\pm 3.2$ and C$^{6+}$/O$^{7+}$ = 2.1 $\pm 0.7$. (ACE data on the abundance of  O$^{8+}$ and the Nitrogen ions was not available during the time of the observations.)  From the observations (Table~\ref{ION}), we derived relative abundance C$^{6+}$/O$^{7+}$ = 2.1 $\pm$ 1. This result of the spectral analysis are in excellent agreement with the ACE measurements.

The obtained relative abundance we derive for C$^{5+}$ from the spectral analysis however, is off by an order of magnitude --  C$^{5+}$/O$^{7+}$ = 87 $\pm$ 29. %As was concluded by Bodewits \e (2007), 
There are several factors that could account for this.  
%First, there may be a contribution to the \CV\ line from other ions in the 250-300~eV range
%(e.g. Si, Mg, Ne) that are currently not included in the model. 
%Secondly, including these species in the model would lower the \CV\ flux, but probably only with a small amount.  
The first possibility is an error in the ionic cross--sections. 
Recent lab experiments suggest that at velocities around 450 km/s, the theoretical emission cross sections for collisions with atomic hydrogen are about a factor of 2
 smaller than measured emission cross sections for collisions between C$^{5+}$ and H$_2$O molecules (Bodewits \e 2007b). 
% Nor, do we believe this is a problem with the calibration. 
 Secondly,  while calibration errors for $Chandra$ at 300~eV are relatively high, cross-calibration with XMM indicates that these are much less than 50\%. 
The quoted error for the absolute effective area of ACIS ``$<5\%$''\footnote{http://cxc.harvard.edu/cal/}. In addition, we see that this has been a persistent effect. Bodewits \e (2007) use O$8^+$/O$7^+$ vs. ionization freeze--in temperature as a measure for the state of the wind. They show that the correlation between ionization freeze--in temperature and C$^{6+}$/C$^{5+}$ is poorly correlated, ostensibly because C$5^+$ abundances are consistently too large. In the case of 17P/Holmes, Christian \e ($in~prep.$) noted a similar issue.

 It is therefore very likely that the C$^{5+}$ emission feature is strongly contaminated by emission from several ions of species such as Mg, Si, Ne, etc (Sasseen \e 2006).
%Fluxes are calculated from 0.31-1.00 keV. This is slightly different than the 0.3-1.0 keV range we have used previously (cf. Lisse \e 2001, Lisse \e 2005, Bodewits \e 2007, etc.).
%During our analysis of 73P/B it was noted that Carbon flux measurements using limits from 0.31 to 1.0 keV have significantly less flux than 0.3 to 1.0 keV. Further, for Holmes the 0.31 to 1.0 keV fluxes were a much better match to contemporaneous observations by $Swift.$ The effect is more pronounced in 17P/Holmes than in the sample of ``harder'' comets discussed by  Bodewits \e 2007. 
The source of the discrepancy is probably related to the energy of the C$^{5+}$ emission line and other low charge state ions, which lie just below 300~eV.  These lines may be more important in 73P/B than in other comets if the charge state of the solar wind is relatively low. %\CV\ is the brightest of all lines in the model. 
The flux is computed from the theoretical spectrum assuming no contribution from lines below 300~eV.
Based on the ACE data, we estimate that as much of 90\% of the emission around 300eV should be attributed to species other than C$^{5+}$. Early Suzaku results show strong emission at low energies, due to a whole forest of unresolved lines of those species (G.V. Brown \e, private communication).

\skipthis{
If the flux measured here and in the other comets is right, then it is
possible that the cross section are wrong.  Laboratory measurements of
C5$^+$ cross sections were done by Bodewits (2007b).  While a small
difference ($<2\times$) was found between the laboratory measurements using
H$_2$O versus H as a target, %estimates which went into the flux
			     %determinations, 
the difference was still smaller than the order of magnitude difference
between the models and the actual cross-sections required to
explain the flux ratios.
}

We performed an experiment using theoretical line strengths from a model under development by Koutroumpa \e (2008).  In this experiment, we extended the model down to 250~eV and added Mg~{\small VIII-X} lines.  The results of this had 
minimal effect on the measured \CV\ flux but indicated a missing low energy component. Following this, we performed a similar experiment using Si~{\small VIII-X} lines.  The effect was dominated by Si~{\small IX} (263~eV) and demonstrated a 25\% reduction in measured \CV\ flux while matching the observed low energy distribution. In a third experiment, we held the carbon \CV\ line fixed at a flux level consistent with the observed solar wind and allowed the Mg lines to vary.  The resultant $\chi^2$ of the new fit was poorer, but by less than 1$\sigma$ than our nominal fit.
We obtained a similar result using the low-lying excitation lines of Mg.
%The results of the latter two experiments are shown in Figure~\ref{EXP}.
This indicates several factors may be in play to produce the relatively high measured flux of the \CV\ line. These factors include the existence of poorly understood lines in this regime, the relatively low effective area below the Carbon edge, and the relatively poor calibration of $Chandra$ below 275~eV.  However, there are neither enough data nor leverage below 310~eV to reliably distinguish \CV\ from Si~{\small VIII-X}, Mg~{\small VIII-X} or any other moderately strong ionic feature.

\subsubsection{Spectral  Morphology}

%DB 22/10/07 - revised

The defining characteristics of the global 
X-ray spectra of a comet are the velocity and state of the solar wind as well as the collisional opacity thickness of the comet. The latter of these depends on $Q_{gas}$ and drops as the distance from the comet cubed. It is not possible to tell solely from the global ionic ratios whether a comet was collisionally thick or thin, simply because elemental and ionic abundances in the solar wind are highly variable and the initial wind conditions could have been anything.
  
We can compare the results of the spectra in the shells of increasing radii using the CXE model (Bodewits \e 2006, 2007). The interaction model assumes solar wind ions penetrating a neutral coma described by the Haser model. The Haser model assumes that a comet with a production rate $Q$ has a spherically expanding neutral coma (Haser 1957, Festou 1981). The lifetime of neutrals in the solar radiation field varies greatly amongst species typical for cometary atmospheres (Huebner 1992). The dissociation and ionization scale lengths also depend on absolute UV fluxes, and therefore on the distance to the Sun. The coma is assumed to interact with solar wind ions, penetrating from the Sunward side following straight--line trajectories. The charge exchange processes between solar wind ions and coma neutrals are explicitly followed both in the change of the ionization state of the solar wind ions and in the relaxation cascade of the excited ions.
A 3D integration, assuming cylindrical symmetry around the comet--Sun axis, finally yields the absolute intensity of the emission lines. Effects due to the observational geometry (i.e. field of view and phase angle) are included at this step in the model.  There are likely to be other, minor effects, not included in the model. For example, the lifetime against photodissociation also depends strongly on solar activity because of the variation in Ly$\alpha$. At high solar activity, not only are the lifetimes of H$_2$O and OH considerably shortened, the dissociation also leads to a higher proportion of excess energy and thus more high-velocity H.

The charge exchange processes between solar wind ions and coma neutrals are explicitly followed both in the change of the ionization state of the solar wind ions and in the relaxation cascade of the excited ions. Bodewits \e (2006)  calculated the effect on line ratios due to the distance from comet center for $Q_{gas}$ rates of 10$^{28}$ and 10$^{29}$ mol. sec$^{-1}$ and a constant solar wind velocity of 450 km sec$^{-1}$. For the most part they find that the various line ratios are fairly flat for the low $Q_{gas}$ case with stronger monotonic trends expected in the cases of \OVIII/\OVII, \CV/\OVII\ and \CVI/\CV.

In Figure~\ref{trends}, we plot the observed trends for \OVIII/\OVII\ and \CV/\OVII\ as a function of the distance from the peak of the emission. The solid line in these figures are  linear least-squares fits, weighted by the measurement errors. The fits show a  
two-fold decreases in  \OVIII/\OVII, as well as a similar increase in \CV/\OVII\  over a span of about 20,000 km. These results are very close to the predictions given by
Bodewits \e (2007; Figure~7) for a low $Q_{gas}$ comet with a modest (500 $\kmps$) wind. The ratio of \CVI/\CV\ is seen to rise as well, again consistent with the prediction of Bodewits \e due to decreases in the density of the neutral gas further from the comet.

It must be noted that there are several caveats to this analysis.  Errors on our observed ratios are fairly high, especially for small values of the radial distance from the nucleus.  % we find several trends that are significant as the radius increases.  We find a steady decrease in the \CV/\OVII\ ratio from $\sim$ 6,000--25,000~km. % from 37.4-23.9  (mean error $\sim 14.5$) t
%The fractional change is very similar to that shown for the $Q_{gas}$=10$^{28}$ mol. sec$^{-1}$ case shown by Bodewits \e (2007) in their Figure~7.  Similarly the \OVIII/\OVII\ ratio is seen to rise by about 50\%.  Again, this is in basic agreement with the prediction albeit somewhat more extreme. 
Figure~\ref{trends} shows fits to the \OVIII/\OVII\ and \CV/\OVII\ data as a function of distance from the peak of the emission near the center of the triangle.  The  data were fitted with error bars only for the ratios, the errors do not account for the positional issue that the emitting particles vary in there distance from the center and are not all at the mean distance - especially in projection.  Nonetheless, the formal probability that the obtained fit is consistent with the correct model is about 90\%.
Morerover, while errors on our observed ratios are fairly high, especially close to the center of the emission, we note that if adjacent regions are combined, the errors shrink in quadrature and the extremal difference, i.e. the significance of the trend, becomes $\sim 3\sigma$. Further, as discussed in the previous section,  the \CV\ line is highly contaminated by low lying lines.  So the changes observed in this feature may not be entirely due to a relatively strengthening of the \CV\ feature towards the center of the emission, but rather a strengthening of the composite of \CV\ and the lower energy  
Si~{\small VIII-X} and Mg~{\small VIII-X}  features.

In terms of flux,  Ellipse~0
is the region of primary emission and accounts for about 65\% of the total
flux measured in the larger emitting region. 
%The values of the line
%ratios for this region represent a flux weighted average of the
%semi-annular regions within it.   This region has a a \CV/\OVII\ line ratio similar to
%the innermost semi-circular region while the  \OVIII/\OVII\ ratio is akin
%to the adjacent annulus -- Ellipse~1.
%In the full ``bigbox'' region.
The ``triangle'' represents an intensity enhancement within this inner ellipse
accounting for about one-third of its the flux in about one-fifth
of the area.  %This region also  shows  \CV/\OVII\ and  \OVIII/\OVII\ line ratios similar to
%the full ``bigbox'' region, clearly consistent with the value for the
%inner ellipse as a whole.
%However, when we examine the point like emission the  \CV/\OVII\
%ratios are almost double the value of the underlying triangle of
%emission. Similarly the \OVIII/\OVII\ line ratios are much smaller
%for the point sources, except in the nuclear regions.
The two-fold decrease in  \OVIII/\OVII\ as well as the similar increase in \CV/\OVII, may imply
that the triangle of emission is brighter than its surrounding by 20\% due to a slight density enhancement in the local coma neutral population.
%in which the increase in the emission is directly proportional to the increase in the number of neutrals. In the three point--like locations, the data are more consistent with density enhancements of factors of 10-20.  This may arise from the fact that these point--like emissions are %the result of out gassing of individual, unresolved particles associated with the disintegration of the fragment 73P/B. 
The low signal to noise of the data prevents a definitive statement in this regard, however.

\subsection{Optical vs. X-ray Luminosity}

  %With a flux of $8.0\times 10^{-13} \erg\ \psqcm ~\ps$ at a distance of about 0.1 AU from $Chandra$, 73P/B had a total X-ray luminosity of $2.3 \times 10^{13} \erg\ \ps$.  
  In an early inter-comparison among comets observed in the X-ray comparing the $L_x/L_{opt}$ ratio served as an argument that whatever is causing the X-rays must be related to the gas and not to the dust in the cometary coma. For this purpose, Dennerl \e (1997) introduced three bins for the gas-to-dust ratio.  Within this binning, the  $L_x/L_{opt}$ ratio appears correlated with $Q_{gas}/Q_{dust}$. $L_x$ was found to follow $Q_{gas}^{0.67}$ for low  $Q_{gas}$ ($L_{opt} < 10^{19.5}$).  At higher values of  $Q_{gas}$ there was an apparent asymptote, with $L_x$ never exceeding about 10$^{16}$ erg sec\minusone.

There were (and still are) considerable uncertainties in the determination of both quantities (in particular due to the temporal variability of $L_x$), but there was also such a considerable dynamic range (3-4 orders of magnitude) for the ``ROSAT comets'' in  $L_x$ and $L_{opt}$ that such a plot, with just three $Q_{gas}/Q_{dust}$ bins, was justified. %In the meantime, many things have changed which may appear to minimize the usefulness of such a plot:
  We now know that charge exchange is the reason for the cometary emission; there is no need anymore to demonstrate that the gas in the cometary coma is the target for this interaction.
In addition, there have been observed cases where $L_x$ was highly variable, but neither $L_{opt}$ nor $Q_{gas}/Q_{dust}$ varied ( e.g. Tempel~1 pre--impact; Lisse \e 2007 and Encke; Lisse \e 1999).  So it is clear that the connection between $L_x$ and the visually observables
is not {\em always} directly related.   In fact, we know that the type of the solar wind is an independent  parameter that influences $L_x$ independent of the comet (e.g. Bodewits \e 2007).
Finally, comets have been observed by various satellites (with different FOVs, different spectral coverage, different background), which makes a comparison of $L_x$ a non-trivial task.

However, now that the asymptotic behavior has been explained as an collisional opacity effect (Lisse \e 2004, Wegmann \e 2004), such a plot should have renewed usefulness in distinguishing collisionally thin from collisionally thick systems.  Indeed, limiting behaviors are seen in these data. The left hand linear branch in Figure~\ref{LXLOPT} is inhabited by faint, low activity comets and is explained by collisional depth. The basic explanation is as a comet has a larger gas production rate, the area over which it emits X-rays becomes larger, too. The surface brightness becomes lower, but the overall brightness increases roughly linearly with the outgassing rate. 
The turnover at L$_x \sim 10^{16} \lum$ is taken to indicate an 
intrinsic size limit for the neutral coma set by the ionization time for the majority coma species, like H$_2$O, CO, and CO$_2$ ($\le10^6$ sec), and the gas outflow rate
($< 1~ \kmps$). This sets the intrinsic size limit  for X-ray emission from any comet at about 10$^6$ km.  The right side of the plot has more scatter and is occupied by high activity, X-ray bright comets.

To calculate luminosities, we measured fluxes from 0.30 to 1.0 keV using the CXE model with solar wind velocities from Bodewits \e (2007) for each comet given in that paper.  Temple~1 and Encke (2003) lie very close to Encke (1997) a low activity comet.  Q4 (Neat) was very active, and moderately X-ray bright 
($\sim 3.4\times 10^{15}$ ergs sec\minusone), and lies near Hyakutake in the middle right of Figure~\ref{LXLOPT}. As noted previously, the flux from 73P/B was calculated from 0.31 to 1.0 keV to correct for concerns near the Carbon edge of the HRMA mirror.  This has a pronounced effect on a soft comet as was 73P/B, but a much smaller effect ($<10\%$) on the other comets in the sample. 
The total observed luminosity from 73P/B of $2.3 \times 10^{13} \erg\ \ps$as derived in \S~3 is consistent with the luminosity derived from Q$_{gas} \approx 10^{28}$ mol. s\minusone\, as well as the simple models of and Cravens (1997).  This is similar to the Q$_{gas}$ found in \S~4.3.3. For optical luminosities below $10^{19}~\lum$ the observed cometary optical and X-ray luminosities track well: $L_x \sim 10^{-4} L_{opt}$.  Tempel~1 and Encke (2003 perihelion) are very close to Encke (1997 perihelion), implying consistency between ROSAT and $Chandra$ observations.  All are low activity comets, in the CXE collisionally  thin regime.

\subsection{When Comets Disintegrate}

The X--ray morphology of cometary fragment 73P/B motivated us to re--examine observations of the CXO observations of CXE  X-ray emission from LS4 on 01 Aug 2004, after that comet's total breakup.  This emission has been a puzzling mystery for quite some time, as no obvious nucleus was seen in the $Chandra$ images, and simple estimates of the lifetime of gas produced during the breakup on 23 July 2000  indicated that there should have been no gaseous material left from the breakup event by the time of the 1 August 2000 observations.  We argue here that the source of the gas was the remnant debris of the comet that continued to outgas, at least through the 1 August observations.

Comparison of the debris field seen in HST of LS4 after its catastrophic breakup in late July 2000, to the HST imagery of 73P/B in April 2007
(cf. Figure~2 Weaver \e 2001) shows an analogous structure, with a large concentration of fragments near the source nucleus, and a fan of material spreading out as radiation pressure and Keplerian motion interact. We can thus expect that a triangular region of X-ray emission expanding towards the northeast could be seen for LS4 on 1 August 2000 by $Chandra$. %In fact, when an analysis of the $Chandra$ LS4 imagery is made, a concentration of emission in the anti-solar and along the direction of orbital motion sector is found, as is an enhancement in the emission in the northeast quadrant.

%As discussed in \S~2.3, the morphology can be understood in as the
%comet flakes apart non-gravitaional forces.
%While large particles stay in nearly Keplerian orbits, in the small particle limit,  the net acceleration is  
%approximately constant and roughly linear in the anti-solar direction.  
%Particulates in between these extremes move along
%cartiod trajectories filling the moving from the direction of motion
%towards the anti-sun direction.  

 To obtain an X-ray spectrum of LS4 after break up, we analyzed the 1 August 2001 observations of LS4 (proposal No. 01100323). This observation was taken about one week after the comet's break-up and we used this to compare with the extended emission around 73P/B. Sixteen individual observations of LS4 were combined for a total exposure time of 18.6~ks.  There was no obvious crescent-shaped comet in the LS4 S3 chip image.  Rather, the image showed several discrete 'knots' and possible diffuse emission
(Figure~\ref{img:ls4}). The total counts in the 300-1000~eV range on the S3 chip were 8430.  This already indicates a fairly strong signal as there were only 6183 counts in that energy range detected on the S3 chip for 73P/B.  The observation time was 18.6 ks for LS4 compared to 20.6~ks for 73P/B. Hence, LS4 was 50\% brighter, despite being further away and completely disintegrated!

Figure~\ref{img:ls4} shows a highly smoothed image of the LS4 data in the 0.3-1.2 keV energy band.  The Sun line and direction of motion are indicated, as is  the location of a small grouping of $>$ 70 m diameter fragments identified by Weaver \e (2001).  Their HST image is shown as the inset.  
The HST image clearly shows a broad fan of scattered light surrounding the big fragments, most likley due to fine dust (and most likely already gas depleted -- beacuse it is fine dust of $\sim 1 ~\mu$m in size, with an outgassing lifetime of $<$ 1 day (Bockelee-Morvan \e 2001).
% Following the discussion in \S4.1, we expect debris to be scattered, roughly symmetrically about the nucleus. 
As for the fragments of 73P/B in the optical HST images, there appears to be a preference for the large debris to be scattered in the anti--Solar direction (the triangle in Figure~\ref{img:ls4}).  % Here the debris were dominated by much smaller gain--sized particles.  The grains are acted on by solar radiation pressure and solar gravity as the comet flakes apart. Particulates move along cardioid trajectories, ejected on the Sun facing side with velocity exceeding the escape velocity and then moving from the direction of motion towards the anti-Sun direction.  The debris from LS4 were subjected to both solar gravitation and radiation forces.
Note that this comet was 5.5 times further away than 73P/B.  If LS4 had been observed at the distance of 73P/B then the solid debris would cover the entire CCD.  The forces on small ice particles are dominated by Solar radiation pressure and hence driven in the northeast quadrant of Figure~\ref{img:ls4}, trailing the nucleus and away from the solar radiation vector.  %We expect that the debris of LS4 are tumbling, hence there is no preferred direction of outgassing -- in contrast to 73P/B.   Ice sublimates from LS4 debris and escapes in a random direction and then interacts with the solar wind leading to a brightening in this quadrant.

The brightening is quantified in Table~\ref{tab:LS4}.  The nominal nucleus is still the brightest region (0.086$\pm$0.007 counts \asec$^{-2}$). However the northeast quadrant is still quite bright (0.075 $\pm$ 0.001counts \asec$^{-2}$).  This is about 10 $\sigma$ higher than the other three quadrants (average 0.065$\pm$0.001 counts \asec$^{-2}$).  When we examine the photon distribution at energies where we expect very limited CXE signal (between 1200~eV and 2200~eV), we find the northeast quadrant has 0.044$\pm$ 0.001 cnts/\asec$^{-2}$  while the other three quadrants average 0.041 $\pm$ 0.001 counts \asec$^{-2}$.  While this is a
3~$\sigma$ detection of excess it may be accounted for by high lying CXE lines, related to highly charged Fe, Si and Mg ions, as observed during the interaction between comet Ikeya-Zhang and a CME (Boedwits \e, 2007Dennerl \e $in~prep$).  Overall this indicate the excess background in the S3 chip observed at low energy was of astrophysical origin -- i.e. outgassing of cometary debris.

A spectrum was produced between 300-1000 eV in a 225\asec\ radius circular aperture containing about 6500 counts and following the same prescription discussed in \S2.3. Figure~\ref{fig:LS4spec} shows both stronger Carbon and Neon emission by LS4 {\bf after} its disintegration than seen in the extant comet 73P/B.  As discussed in \S4.2, the spectral structure depends primarily on the composition and ionization of the solar wind and secondarily on the density of the gas.  In both cases, the cometary gas was optically thin to the CXE lines and the solar wind speeds are nearly identical ($\sim 450~\kmps$ for LS4). However LS4 was observed during solar maximum and the more highly ionized species, \OVII\ and especially \OVIII\ dominated the solar wind.  Hence the shape of its spectrum is not too different from the pre-breakup spectrum (Lisse \e 2001).  The key distinction appears to be the solar wind ionization state; LS4 has a ``hot'' spectrum while 73P/B has a colder spectrum. % with some indication of mixing-in of hotter material.

\section{Conclusions}

\begin{figure}
\plotone{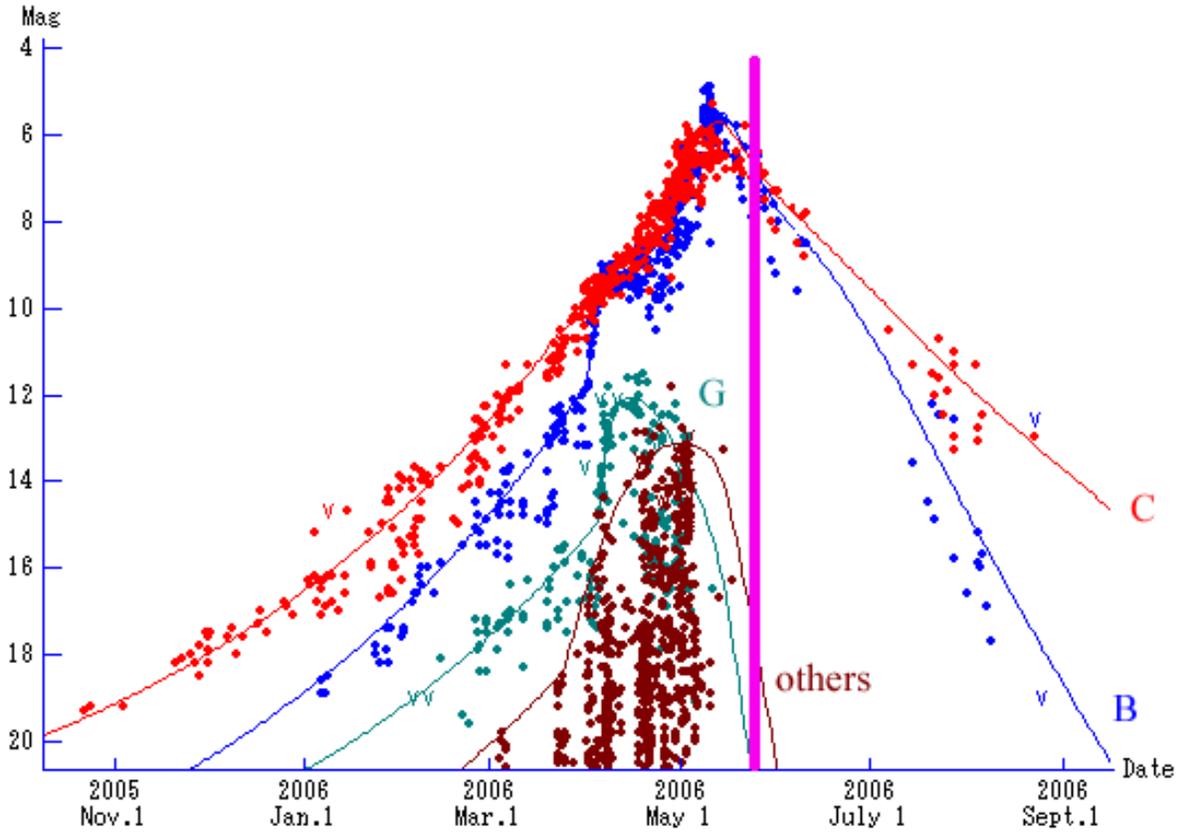}
\caption{Lightcurve of various fragments of comet
  73P/Schwassmann-Wachmann 3.
Compiled by Yoshida  (2007; http://www.aerith.net/comet/catalog/0073P/index.html.)
Fragment B brightened suddenly in April 2006 and was as optically bright as the
  C Fragment at the time of the $Chandra$ observation -- indicated by the vertical
  line.}
\label{lightcurve}
\end{figure}

\begin{figure}
\plottwo{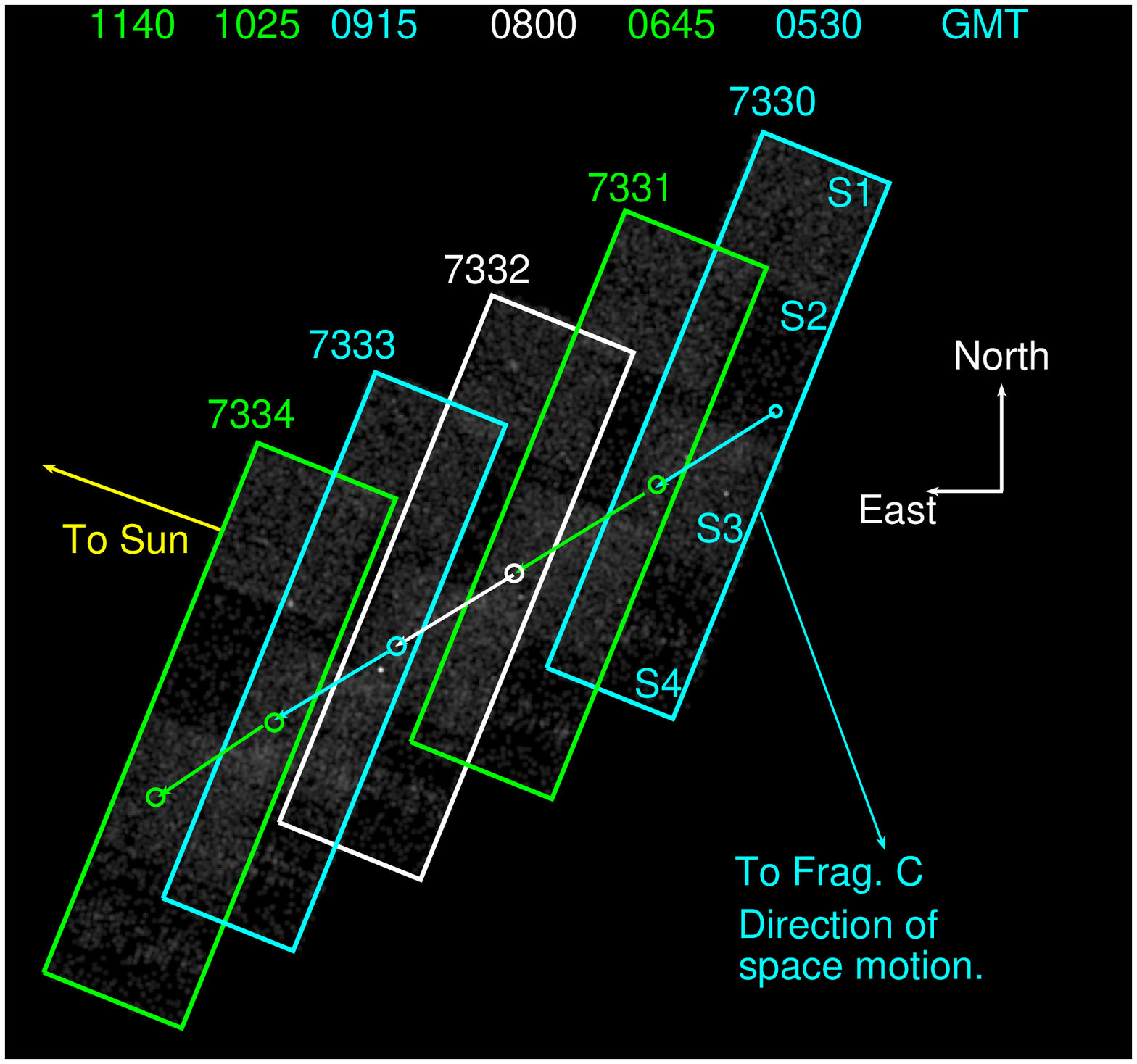}{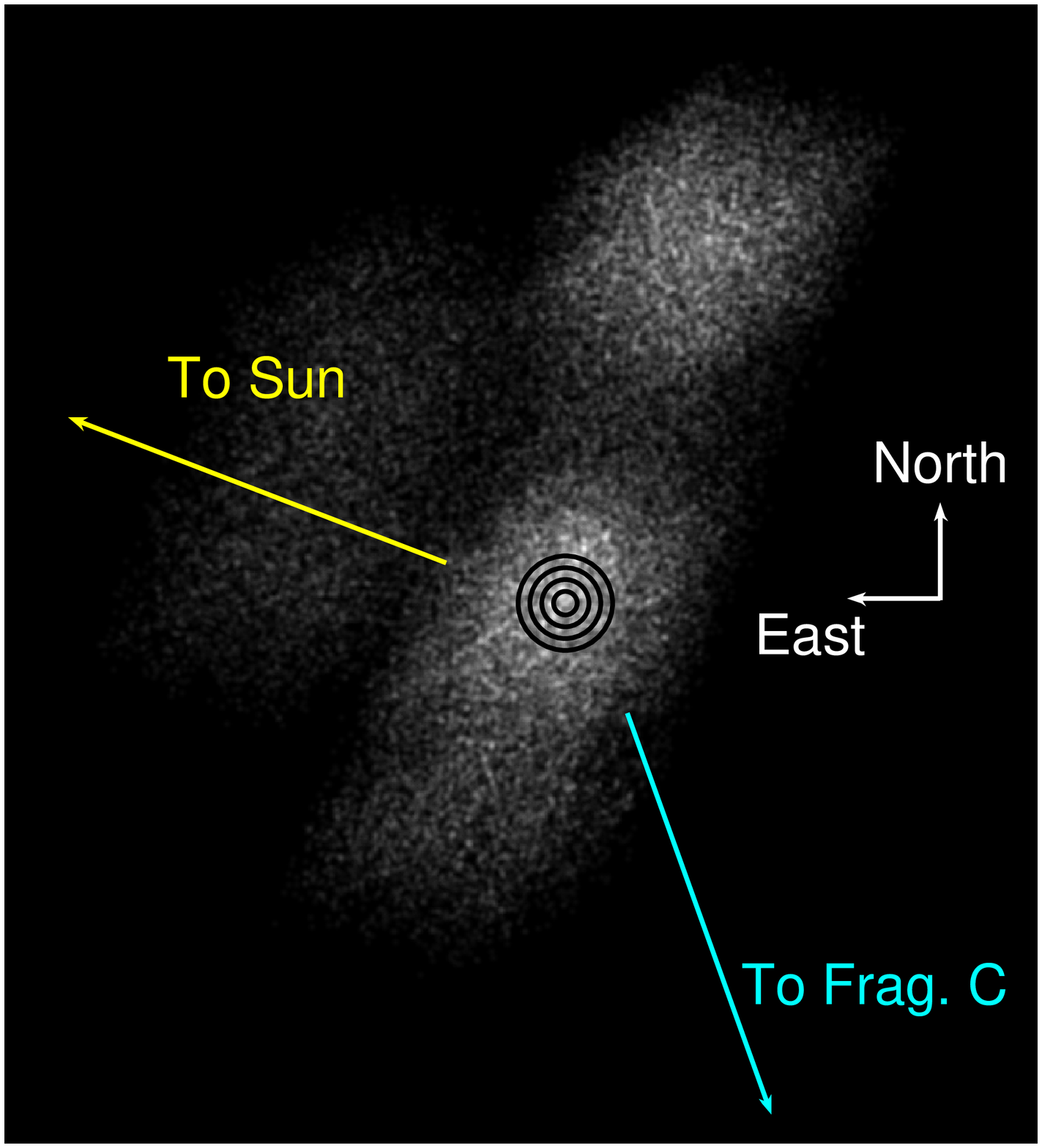}
\caption{Layout for $Chandra$ observations.  \textbf{Left:} In the frame of the spacecraft. The S1-S4 chips are rendered from all 5 pointings (Obsids 7330-7334).  Chip IDs are noted in observation 7330.  73P/B tracks about 140 degrees east of north.  This is mostly reflex motion, the actual space motion of the comet toward Fragment C, about 210 degrees east of north.
The Sun is 70 degrees east of north.  The numbers along the top map to circles along the track and indicate the time the comet was at a particular position.  All the times indicate the beginning of an observation -- in UT 23 May 2006 -- except ``1140" which was the end of the last observation. \textbf{Right:} The same data, with chips I2 and I3 included and all data reprojected onto the comet's frame of reference. The ``bulls-eye" pattern indicates the location of the nucleus.  The apparent signal in the S1 chip at the upper right is not due to X--ray emission from 73P/B. The diameter of the rings are 1\arcmin\ - 4\arcmin\ inclusive.
\label{LAYOUT}}
\end{figure}

\begin{figure}
    \centering
        \includegraphics[width=\textwidth]{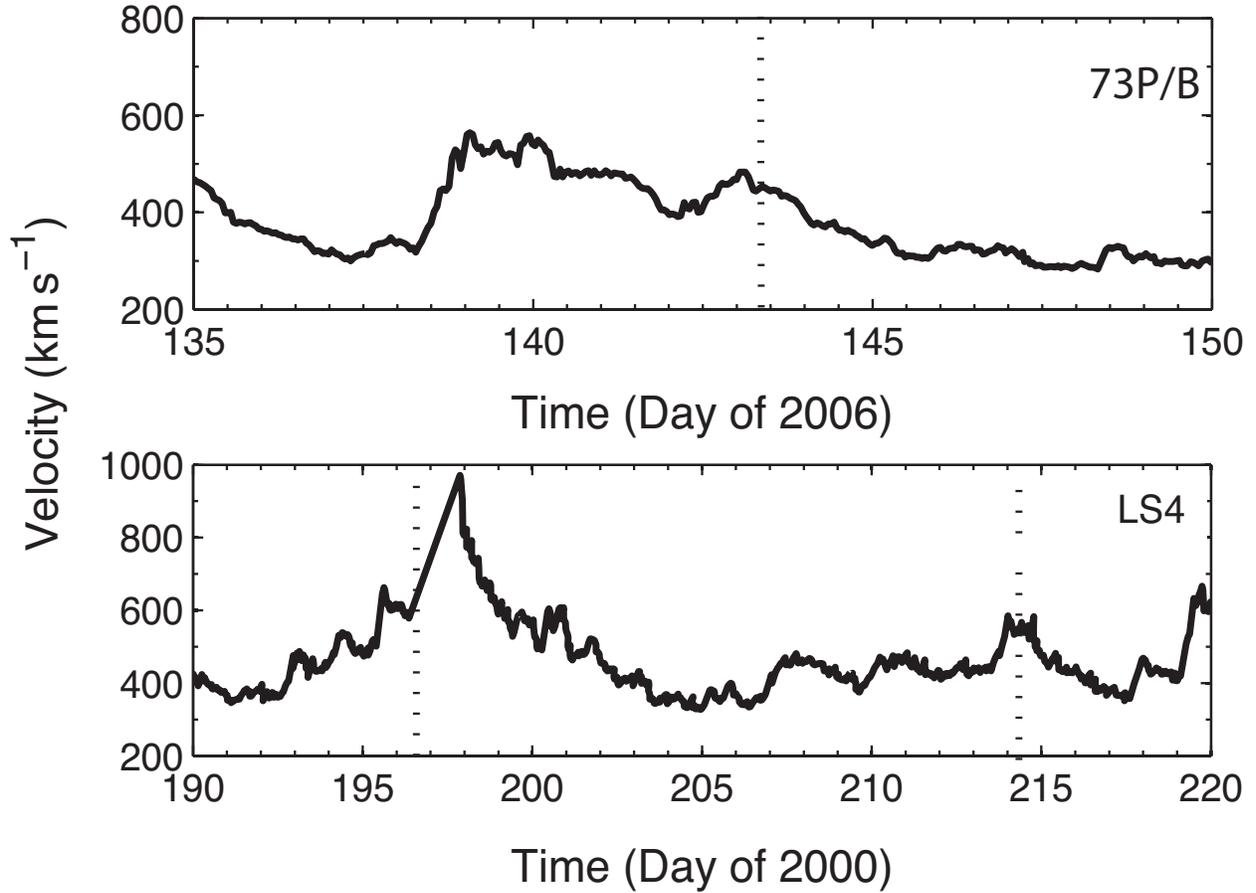}
    \caption{Solar wind proton velocities estimated from SOHO proton
    monitor data for 73P/B \textbf{(top)} and LS4 \textbf{(bottom)}. The time of the observations is indicated with a
    dotted line. The data were mapped to the position of the comet using a radial and
    co-rotational extrapolation scheme.
\label{fig:sw}}
\end{figure}

\begin{figure}
    \centering
       \plotone{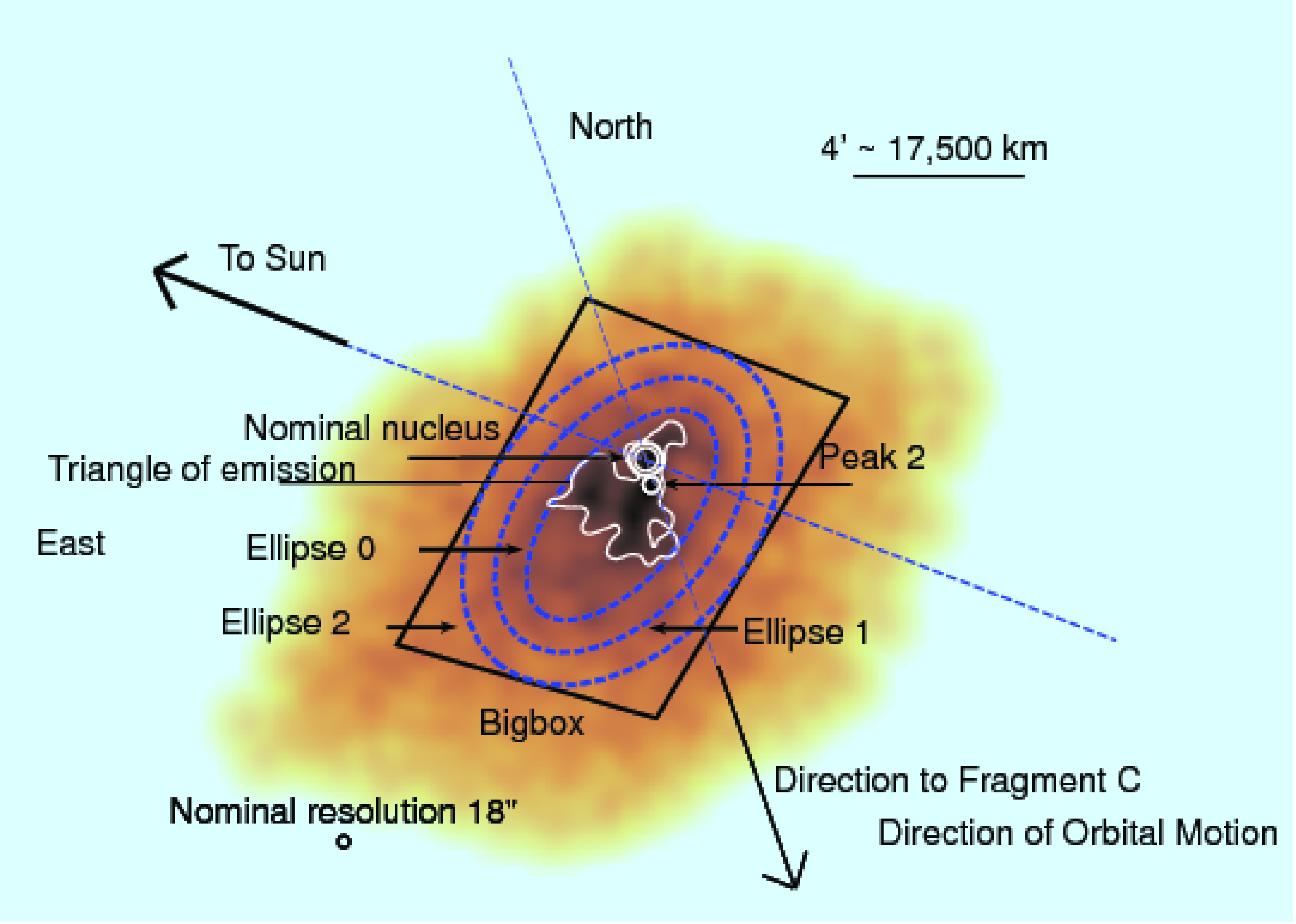}
       %\plotone{SW3_R_PL12_nohigh_G_lin12_nohigh_B_Log_invert_markup.ps}
           \caption{$Chandra$ image of 73P/B. Data from CCD S3 are shown.
     Image has been scaled and
    stretched to highlight details.  Red has been power law
    stretched to highlight bright, high spatial frequency ($<$18\asec)
    features. Green has a less extreme squared stretch to
    highlight moderate intensity, high spatial frequency features. Blue shows a
    linear stretch to show the extent of all X-rays.  Colors are
    inverted to enhance contrast (for presentation purposes).  Thus high frequency intense features are 
    black, moderate features are red to orange and faint diffuse features are yellow.
    Regions of interest are identified, including the elliptical
     emission
     regions (blue dashed lines), the triangle of emission  (solid white contour), various peaks within the
      triangle of emission includng the nucleus (solid white circles).  The effective PSF,
     critical vectors and a length
     scale bar are also shown as are additional regions used for spectral extraction and tabulated 
     in Tables 2. Bigbox, Ellipse~0 and Circ100 refer to all data within the region.
      Ellipse~1 and Ellipse~2 are annular regions excluding all data outside of the 50\asec\ wide ring.
      Likewise, Ellinner is Ellipse~0 with Circ100 excluded.  The image is not corrected for exposure     
      variations.
\label{fig:image}}
\end{figure}

On 23 May 2006, we observed comet 73P/2006 (Schwassmann--Wachmann 3 -- Fragment B) for about 20 ks.  At that point, the comet was closer to the Earth than any comet yet detected in X-rays.   73P/B was  0.97 AU from the Sun and 0.1 AU from the
Earth, with a total visual magnitude of V $\sim$ 7.5 (Figure~\ref{lightcurve}), a total luminosity of L$_{optical} \sim  7.25 \times 10^{17}$ erg sec$^{-1}$.  At the time of the observation, the fragment was relatively active, approximately the same brightness as the primary ``C'' fragment.  The main difference between this comet and previous comets observed by $Chandra$ was that its geocentric distance at the time of observation was extremely small, allowing for nearly in-situ measurements of the solar wind by near--Earth satellites.
The $Chandra$ observations, folded through the models of Bodewits \e (2007b) were consistent with the C$^{6+}$/O$^{7+}$ ratio observed by ACE/SWICS.
However, the C$^{5+}$/O$^{7+}$ratios were highly discrepant. Since we cannot attribute the discrepancy to deficiencies in our knowledge of the C$^{5+}$ cross sections, the $Chandra$ calibration, or differences in the local solar wind, we are left to conclude that numerous additional species close in energy (200 - 300 eV) to \CV\ have contaminated our spectral analysis. Future observations with a microcalorimeter are required in order to get a satisfactory understanding of this problem.  In addition, the results of this work show:
%L$_{X} \sim  3.35 \times 10^{13}$ erg sec$^{-1}$.

\begin{itemize}

 \item The X-ray count rate from 73P/B was very stable -- within 10\% of 0.0045$\pm 0.0015$ counts per second.

\item The overall morphology of the emission is roughly elliptical, with the triangular concentration within the ellipse of emission.  One side of the triangle is aligned with the direction of orbital motion; the other is aligned with the Sun.  This morphology can be understood as symmetric, emission dominated by several large pieces of 73P/B  moving close to the comet's orbital path.  The composite of these fills the triangle with Sunward emission.

 \item The X-ray flux from the entire comet ``bigbox" between 0.31 -- 1.0 keV was about  $8.0\times 10^{-13} \flux$.  This is a luminosity of $2.3\times 10^{13} \lum$.  About 25\% of the flux comes from a small triangular region located forward of the nucleus relative to the direction of motion.   This region is about 20,000 km on a side.  % From Table 2.

 \item The spectrum of the comet is well fitted by a charge exchange model containing four species each of two ions -- C, N, O and Ne --  and the observed solar wind speed of about 450 $\kmps$.  Comparison of the line fluxes with the ionic ratios in the solar wind raise concerns that the flux attributed to \CV\ may not be entirely due to \CV. Lines lying below the nominal 300~eV cutoff of the HRMA+ACIS system may be contributing to the measured flux.  We are encouraged that work on identifying possible contributors to these lines is being actively pursued. 

\item Following Bodewits \e (2007), comet 73P/B interacted with the cold--fast solar wind.  This is a reflection of the solar wind being relatively fast, but the ionization fraction in the solar wind of \OVIII/\OVII\ was very low -- the lowest observed in the $Chandra$ era -- despite some evidence of mixing with the CIR.\footnote{Comet 17P/Holmes, observed in November 2007 may have been immersed in an even colder solar wind.  These data are still being analyzed.}  The cold ionization state means that ionized Oxygen in the solar wind is preferentially \OVI\, not \OVII\ or \OVIII. This is one reason the X-ray luminosity is low -- the CXE driven emission from \OVI\  maybe primarily in the EUV. Comet 73P/B is, spectrally, very similar to the observation of comet 2P/Encke 2003.
%Hence, LS4 post-breakup showed stronger \CVI, \OVIII\ and \NeIX\ than 73P/B. 

   \item We analyzed the spectrum as a series of elliptical annuli and found the line ratios of \OVIII/\OVII\ (653/561~eV) and \CVI/\CV\ (367/299~eV) decrease  as one moves from the outer regions of the comet towards the peaks of the emission.  Meanwhile \CV/\OVII\ (299/561~eV) increases as one moves from the outside in.  While the individual measurements have $\sim 30\%$ errors, the trend is significant beyond the 3$\sigma$ level.  This is confirmation of the predictions of Bodewits \e (2007) of the behavior of emission ratios within a collisionally thin medium undergoing charge exchange.

\item Despite its ongoing break up, 73P/B's luminosity is similar to other low activity ecliptic comets with
  $L_X/L_{opt} \sim 10^{-4}.$ 

 \item  The lower limit to the outgassing rate as determined by the observed flux intensity and comparison to Tempel~1/Deep Impact is at least $3 \times 10^{27}$ mol.~sec$^{-1}$. The true rate is probably a factor of a few higher, but less than the $2  \times 10^{28}$ mol.~sec$^{-1}$ observed by Schleicher using narrow band photometry a week earlier.

 \item Extending the morphological analysis we can also understand the observation of LS4 in August of 2000.   Smaller pieces of debris for this comet, influenced much more strongly by solar radiation pressure, caused the LS4 debris to preferentially occupy the region northeast of the nominal position of the (former) nucleus.
This would have been classified as a hot--fast solar wind comet.
Hence, LS4 post-breakup showed stronger \CVI, \OVIII\ and \NeIX\ than 73P/B.

\end{itemize}

\acknowledgments
We thank Harvey Tananbaum for use of Director's Discretionary time for this unusual object.  We thank Dimitra Koutroumpa \& Vasili Kharchenko for helpful discussions on the \CV\ contamination issue. We are grateful for the cometary ephemerides of D.~K.~Yeomans published at the JPL/Horizons website.
Proton velocities and ion measurements used here are courtesy of the
SOHO/CELIAS/PM and ACE/SWICS teams, respectively.  The CXC guest investigator program supported this work through grants DD6-7040X (SJW) and GO4-5167X (CML).  SJW was supported by NASA contract NAS8-03060. DB wishes to acknowledge travel support from the Netherlands Organisation for Scientific Research (NWO), the High Energy Astrophysics Division at the Center of Astrophysics, and the Visiting Scholars Program of the Insititute of Theoretical Atomic, Molecular and Optical Physics at the Center for Astrophysics.

%%%%%%%%%%%%%%%%%%%%%%%%%%%%%
%%%%%%%T A B L E S %%%%%%%%%%%%%%%
%%%%%%%%%%%%%%%%%%%%%%%%%%%%

\begin{deluxetable}{lrrrrr}
%\tabletypesize{\small}
\tablecaption{Total Counts in Various Regions of 73P/B. \label{SurfBright}}
\tablewidth{0pt}
\tablehead{
\colhead{Location} &\colhead{Area} & \colhead{Counts$^1$ }
 & \colhead{Surface}&\colhead{Error} &  \colhead{Sigmas}
 \\
\colhead{~}  & \colhead{(pixels)} & \colhead{(net)} &
\colhead{Brightness} & \colhead{Surf. bri.} &
\colhead{above}
 \\
\colhead{~}  & \colhead{} & \colhead{} &
\colhead{cnts/pix.$^2$ } & \colhead{cnts/pix.$^3$ } &
\colhead{Background}
}
\startdata
All events on I2    &1048580 &  1148  & 1.09 &3.23& --\\
All events on I3    &1048580 &  1005  & 0.96 &3.02&--\\
All events on S1    &1048580 &  6115   & 5.90 &7.45& --\\
All events on S2    &1048580 &  1472   & 1.40 &3.65& --\\
All events on S3    &1048580 &  6183  & 5.89 &7.50 & --\\
All events on S4    &1048580 &  1598  & 1.52 &3.81 &--\\
Background (on S3)         & 384845 &   1121   & 2.91 &9.18 & --\\ %******
Big Box             & 922907          &  4673 & 5.06 &7.41  & 29\\
Ellipse~0             &256072 &   1702   & 6.65 &16.11 &  23\\
Ellipse~1 (-E0)    &210801 &  1149    & 5.45 &16.08 &  16\\
Ellipse~2 (-E1)    &273633 &   990  & 3.62 &11.50 &  6\\
Triangle            & 64335.7 &    783   & 12.17 &43.49  &  20\\
Nucleus (circ100) & 31415.9 &    279   &  8.88 &53.16   &   ---\\
Peak 2  &  2259.8 &     30              & 13.2 &242.24 &   $<3^4$\\
%Peak 3  &  1963.5 &     26               & 13.24 &259.69  &  $<3^4$\\
%peak 3  &  1868.8 &     25               & 13.38 &267.55 &   $<1^4$\\
%little peak to the south&  1728.4 & 23   & 13.31 &277.47  &   36\\
\enddata
\tablenotetext{1}{Energy range is 0.3-1.0 keV}
\tablenotetext{2}{$\times 10^{-3}$}
\tablenotetext{3}{$\times 10^{-5}$}
%\tablenotetext{3}{Energy range is 0.5-1.2 keV}
\tablenotetext{4}{Above the ``triangle''}

\end{deluxetable}

\begin{deluxetable}{lrrrrrrrr}
\rotate
%\tabletypesize{\scriptsize}
\tabletypesize{\footnotesize}
%\tabletypesize{\small}
\tablecaption{Results of the  v=450 km sec$^{-1}$ CXE model fit for the 
73P/B spectra sorted
by spatially selected region. Fluxes are given in units of
ergs cm$^{-2}$ sec$^{-1}$. Errors
are obtained by averaging over the calculated $\pm$ 90\% confidence
contours (corresponding to $\chi^2$ =2.7 or 1.6 $\sigma$). }
\label{tab:spectral}
\tablewidth{0pt}
\tablehead{
\colhead{Energy}&\colhead{Circ100} & \colhead{EllInner$^a$}
& \colhead{Ellipse~0} &\colhead{Ellipse~1} & \colhead{Ellipse~2}&\colhead{Bigbox}
& \colhead{Triangle} &\colhead{LS4(part2)}
}
\startdata
653~(\OVIII)     & $<$0.1$^b$      & 0.7 $\pm$ 0.2   & 1.4 $\pm$ 0.4  & 1.2 $\pm$ 0.3   & 1.3 $\pm$ 0.3  & 3.4 $\pm$ 0.9  & 0.6$\pm$0.2   & 10.1 $\pm$ 1.2  \\	
561~(\OVII)      & 1.3 $\pm$ 0.3   & 3.6 $\pm$ 0.8   & 7.1 $\pm$ 1.0  & 4.4 $\pm$ 0.6   & 4.3 $\pm$ 0.6  & 14.9 $\pm$ 1.2 & 3.2$\pm$0.6   & 8.4 $\pm$ 1.6   \\	
367~(\CVI)       & $<$2            & $<$13           & 11.9 $\pm$ 7.0 & 9.6 $\pm$ 5.4   & 7.5 $\pm$ 5.4  & $<$25          & $<$6          & 33 $\pm$ 8     \\	
299~(\CV)        & 57.3 $\pm$ 16.4 & 134 $\pm$ 63    & 234 $\pm$ 72   & 140 $\pm$ 61    & 102 $\pm$ 52   & 518 $\pm$ 169  & 112$\pm$35    & 65 $\pm$ 33   \\	
500~(\NVII)      & $<$0.6          & $<$1.1          & $<$1.5         & $<$0.8          & $<$0.7         & $<$2           & $<$1.0        & $<$4          \\	
419~(\NVI)       & 1.4 $\pm$ 0.8   & $<$5            & 9.2 $\pm$ 4.3  & 5.2 $\pm$  3.2  & $<$4           & 19.5 $\pm$ 8.9 & 6.5 $\pm$ 2.3 & $<$8          \\	
907~(\NeVIII)    & $<$0.1          & 0.3 $\pm$ 0.1   & 0.5 $\pm$ 0.1  & 0.4 $\pm$ 0.1   & 0.3 $\pm$ 0.1  & 1.1 $\pm$ 0.5  & $<$0.1        & 5.6 $\pm$ 0.9  \\	
1023~(\NeIX)     & $<$0.1          & $<$0.1          & 0.3 $\pm$ 0.2  & 0.3 $\pm$ 0.1   & 0.4 $\pm$ 0.2  & 1.3 $\pm$ 0.6  & 0.2$\pm$0.1   & 5.6 $\pm$ 1.1  \\	
\hline													 		    							
$\chi^2$         &52.1             &  56.3           &61.4	       &   38.2   	 &   40.5 	 & 41.0            &	46.4	  &   96.8		           \\	
$\chi^2$/dof     & 1.27            & 1.37            & 1.50	       &   0.95          &   1.04 	 & 1.0             &	1.13	  &   2.36		            \\	
\hline													                  
Log model flux   &   -13.21        &    -12.63       &    -12.37      &   -12.56        &   -12.65       &    -12.10      &    -12.73     &   -12.14\\				
\hline													 		    							
Line Ratios \\												 		    							
653/561          &  0.0            & 0.18 $\pm$ 0.07 & 0.19 $\pm$ 0.06& 0.26  $\pm$ 0.07& 0.30 $\pm$ 0.08&0.23 $\pm$ 0.06 &0.20 $\pm$ 0.08 &1.20  $\pm$ 0.27  \\			 
367/299          &  0.01           &  0.06           & 0.05 $\pm$ 0.03&0.07 $\pm$ 0.05  &  0.07$\pm$ 0.06& 0.03           &   0.02         & 0.52   $\pm$ 0.26       \\			 
299/561          & 16.3 $\pm$ 6.0  & 37.4 $\pm$ 19.4 & 33.0 $\pm$ 11.1& 32.1 $\pm$ 14.7 & 23.9 $\pm$ 12.6& 34.7 $\pm$ 11.7& 35.2$\pm$ 12.8 & 7.7 $\pm$4.2          \\
\enddata
\tablenotetext{a}{ Ellipse~0 with ``Circ100" emission subtracted.}
\tablenotetext{b}{ ``$<$" denotes a 90\% confidence upper limit}
\end{deluxetable}

\begin{deluxetable}{lrr}

\tablecaption{Solar wind abundances relative to O$^{7+}$, obtained
  from the CXE-model fit. 
\label{ION}}
\tablewidth{0pt}
\tablehead{
\colhead{Ion} &  \colhead{73P/B} &	\colhead{LS4 (post break-up)}}
\startdata
O$^{8+}$   &	 0.18 $\pm$ 0.05  &  0.95 $\pm$ 0.21        \\
C$^{6+}$   &	 $<$2.09          & 4.89 $\pm$ 1.51     \\
C$^{5+}$   &	 87  $\pm$ 29     &  19 $\pm$ 10      \\
N$^{5+}$   &	 $<$0.10          &  $<$0.37      \\
N$^{6+}$   &	 1.76  $\pm$0.82  & $<$1.28       \\
Ne$^{10+}$ &	 0.04 $\pm$ 0.02  &  0.33 $\pm$ 0.09  \\
\hline
\enddata
\end{deluxetable}

\begin{deluxetable}{lrrrr}
%\tabletypesize{\small}
\tablecaption{Total Counts in Various Regions of the LS4 FOV \label{tab:LS4}}
\tablewidth{0pt}
\tablehead{
\colhead{Location} &\colhead{Area} & \colhead{Counts}
 & \colhead{Surface}&\colhead{Error}  \\
\colhead{~}  & \colhead{(arcsec)} & \colhead{~} &
\colhead{Brightness} & \colhead{Surf. bri.}   \\
\colhead{~}  & \colhead{} & \colhead{} &
\colhead{cnts/pix.$^1$} & \colhead{cnts/pix.$^1$}
}
\startdata
Nuclear position$^2$        &  1963.5 &     169   &  21.52 & 1.655\\
Southeast Quadrant$^2$      &  9188.2 &     642   &  17.46 & 0.688\\
Northeast Quadrant$^2$      & 31071.3 &    2327   &  18.72 & 0.388 \\
Northwest Quadrant$^2$      & 15109.2 &     886    & 14.67 & 0.493 \\
Southwest Quadrant$^2$      & 21116.3 &    1399   &  16.56 & 0.443\\
Nuclear position$^3$        &  1963.5 &      80    & 10.19 & 1.140\\
Southeast Quadrant$^3$      &  9188.2 &     353   &   9.60 & 0.510\\
Northeast Quadrant$^3$      & 31071.3 &    1357   &  10.97 & 0.298\\
Northwest Quadrant$^3$      & 15109.2 &     589    &  9.75 & 0.403\\
Southwest Quadrant$^3$      & 21116.3 &     897   &  10.62 & 0.335\\ 
\enddata
\tablenotetext{1}{$\times 10^{-3}$}
\tablenotetext{2}{Energy range of 0.3-1.2 keV}
\tablenotetext{3}{Energy range of 1.2-2.2 keV}
\end{deluxetable}

%%%%%%%%%%%%%%%%%%%%%%%%%%%%
%%%%%%F I G U R E S %%%%%%%%%%%%%%%
%%%%%%%%%%%%%%%%%%%%%%%%%%%%

\skipthis{
\begin{figure}
    \centering
       % \plotfiddle{fig5.ps}{-1.0in}{90}{360.}{500.}{-50.}{-500.}
       \plotone{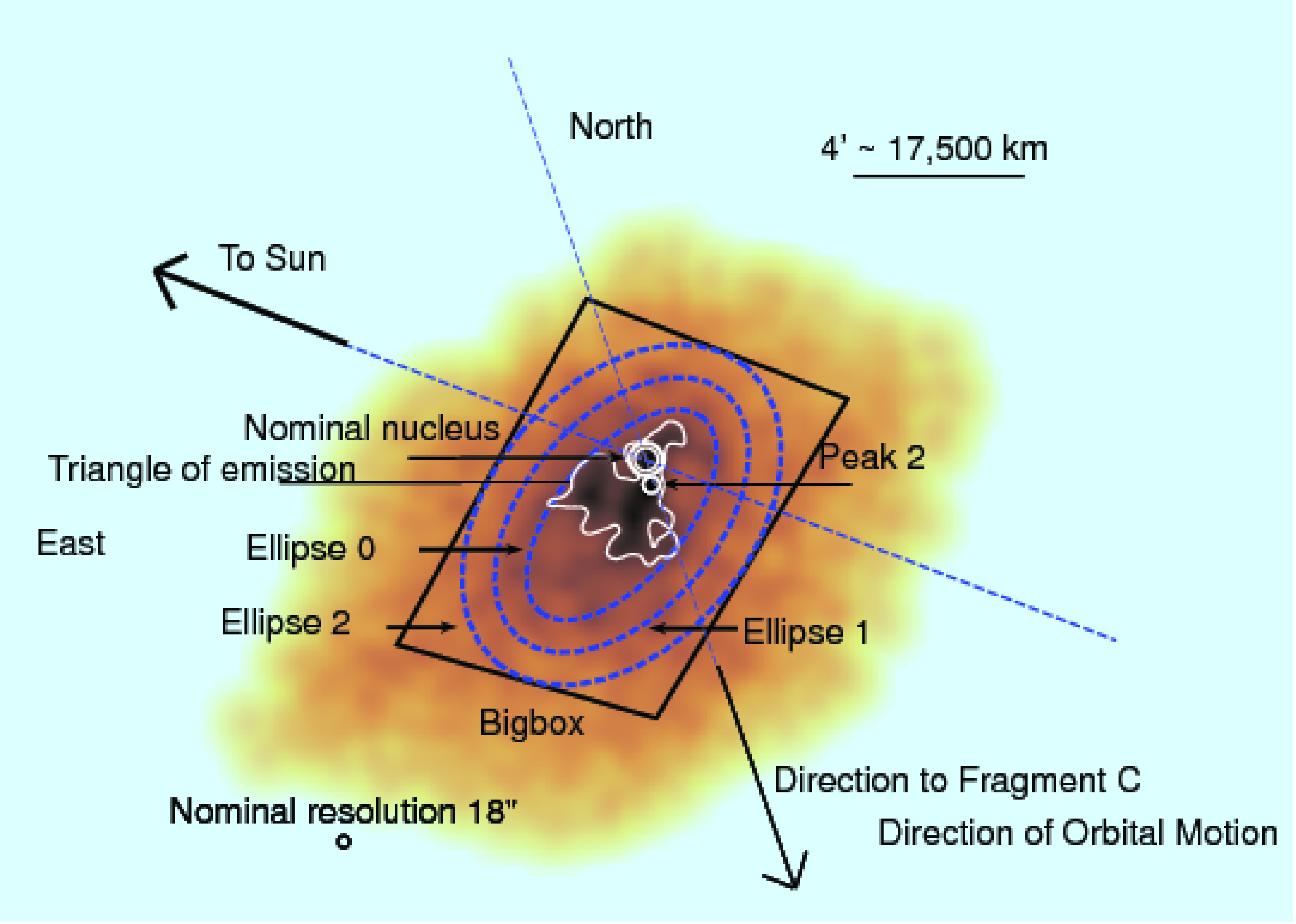}
   \caption{Lightcurve of X-ray events with energies between 0.3~keV
    and 1.2 keV and
    within the triangular region, including the nominal B nucleus.
    The lightcurve is averaged over 200
    seconds and normalized to counts per second.
    The bins near the four data gaps are uncorrected for
    exposure time.  The dot-dashed lines indicate the mean count rate for
    each observation (corrected for exposure time), the solid
    line indicates the mean count rate for the full observation. This
    latter value is 0.045 counts per seconds with a 1 $\sigma$
    deviation of 0.002.
\label{fig:lc}}
\end{figure}
}

\begin{figure}
    \centering
        \plotone{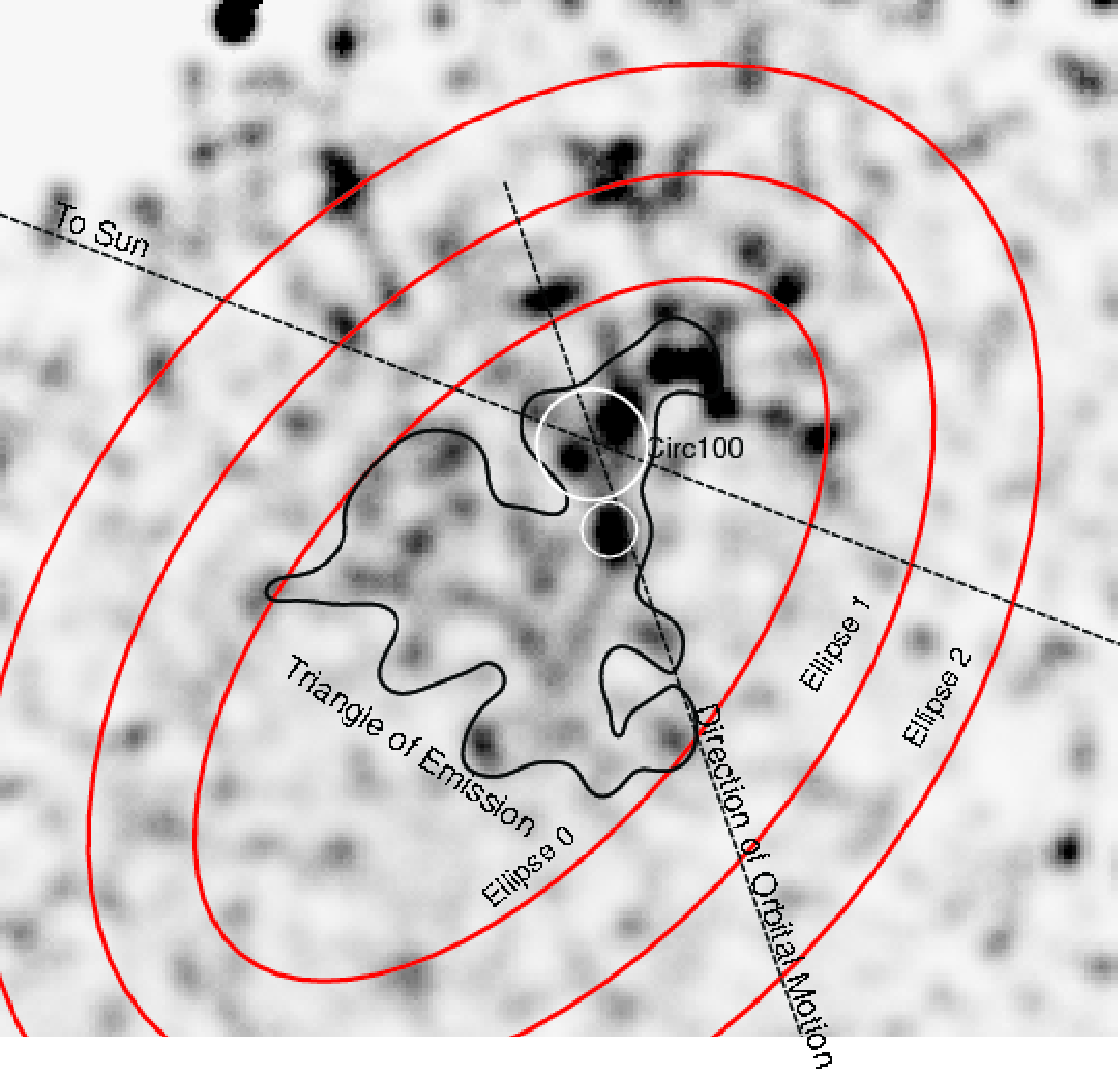}
   \caption{Detail of the central $\sim$20,000~km of the ACIS field.  These
     data have been flux corrected by the exposure time and smoothed
     to about  18\asec.  A similar markup to Figure~5 is used.  
The extreme stretch of this image makes global details less apparent than Figure~5.
The triangle of emission is less pronounced than in the non-flux
     corrected image but still present.
\label{DETAIL}}
\end{figure}

\begin{figure}
    \centering
         \plotfiddle{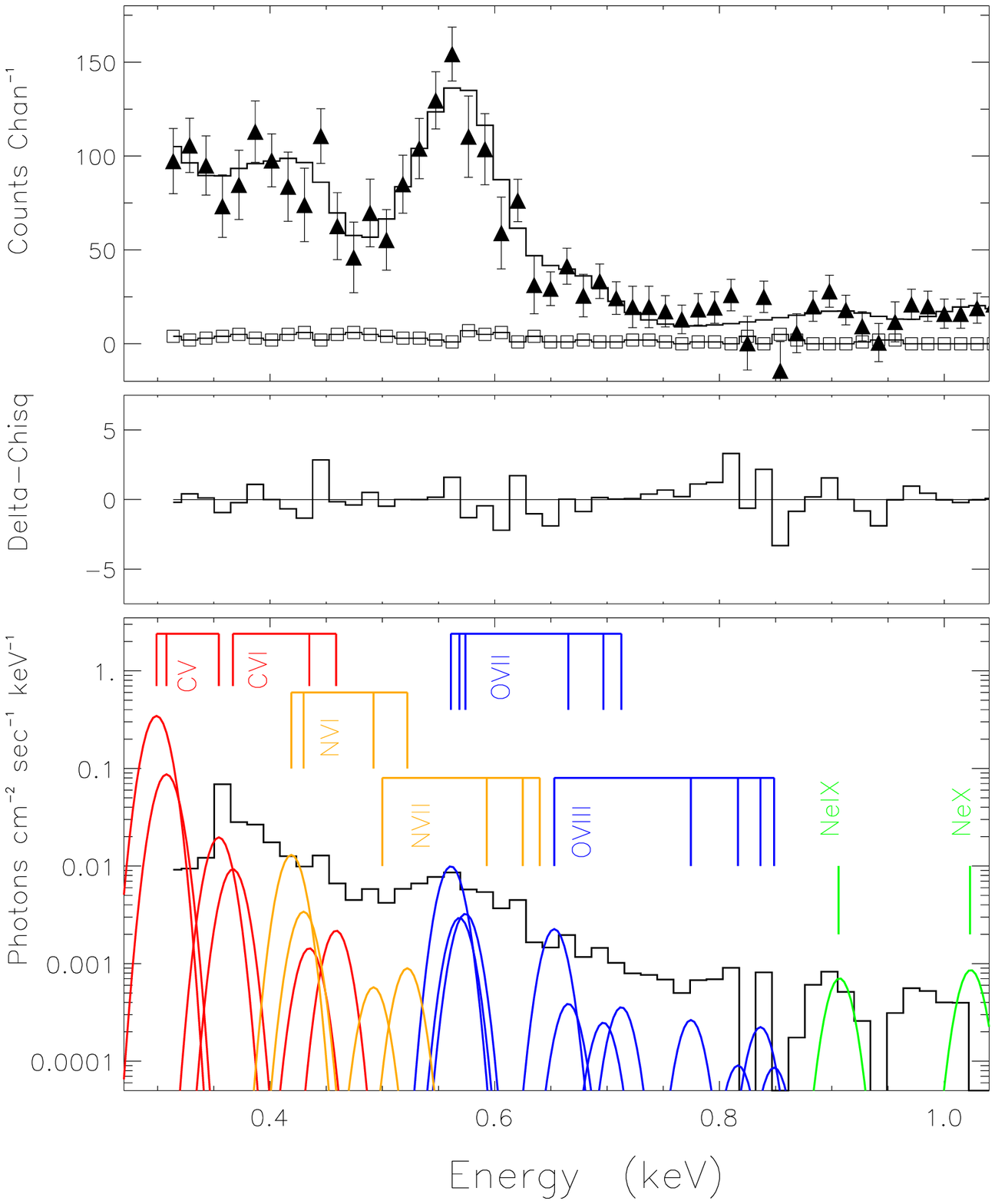}{0.0in}{00}{300.}{400.}{-50.}{-500.}
    \caption{Details of the CXE model fit for the spectrum of comet
    73P/B for the whole "bigbox'' region. \textbf{Top panel:} Shown are the 73P/B count rate 
    spectrum (filled triangles), a sample background (open squares), and best fit
    spectrum with the CXE model (solid histogram). \textbf{Middle
    panel:} Residuals of the CXE fit ($\chi^2 /d.o.f = 1.0$). \textbf{Bottom
    panel:} CXE model and observed spectrum
    indicating the different lines in the fit and
    their strengths. Carbon - red; Nitrogen - orange; Oxygen -
    blue; Neon - green. The unfolded model is scaled above the
    emission lines for the ease of presentation.   The apparently very strong \CV\
    line at 300~eV is enhanced due to contributions from lines at 250 -- 300~eV.
\label{fig:3specstuff}}
\end{figure}

%%%%%%%%%
%%%%%%%%
\clearpage
\begin{figure}
    \centering
         \plotfiddle{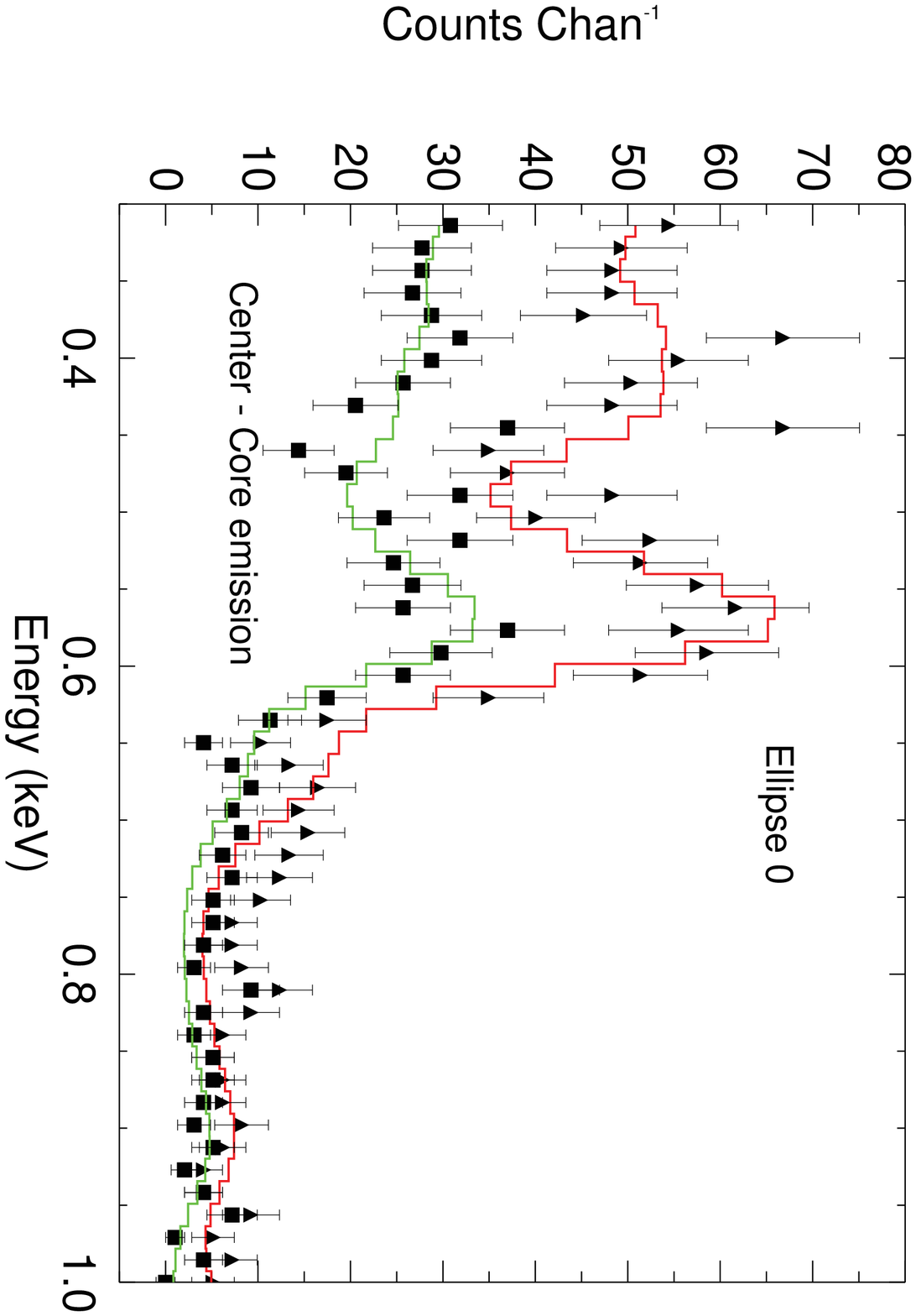}{-1.0in}{90}{360.}{500.}{-50.}{-500.}
    \caption{Comparison of  the spectra near the comet's nucleus. We
compare the central Elliptical region (Ellipse~0) to the %very center
R=100 pixel circle (dubbed ``Center-core") centered on the nominal nucleus.
Data within each region are fitted by a CXE model with a solar wind
velocity of 450 km sec$^{-1}$ and highly ionized species of Oxygen, Carbon,
Nitrogen and Neon. \label{fig:specpoly}}
\end{figure}

%%%%%%%%%
%%%%%%%%
\clearpage
\begin{figure}
    \centering
               \plotfiddle{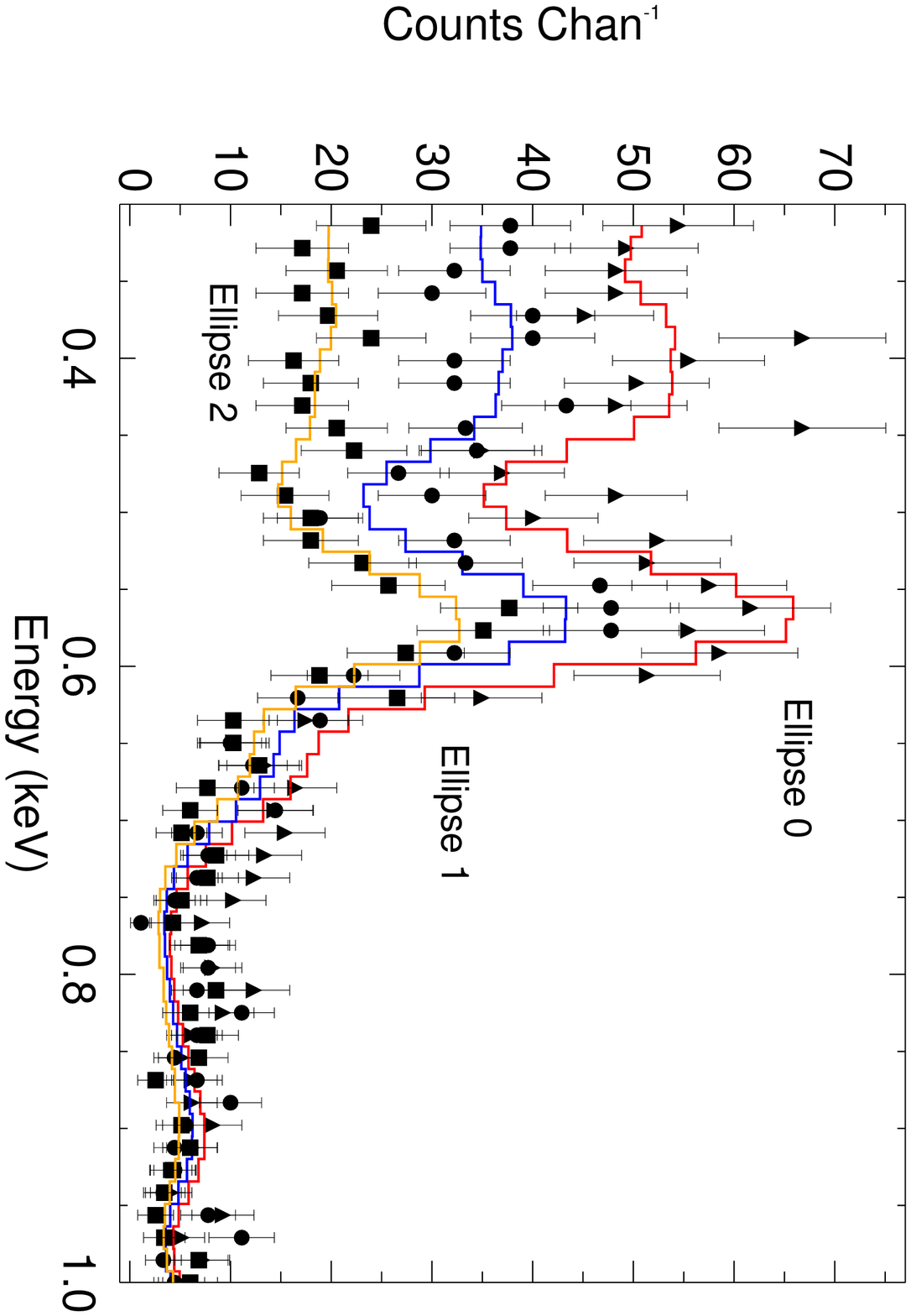}{-1.0in}{90}{360.}{500.}{-50.}{-500.}
    \caption{Comparison of count rate spectra for the larger-scale structure of the comet. We
compare the three elliptical annular apertures with semi-major axis of:
380, 480 and 580 pixels for Ellipse~0, Ellipse~1, and Ellipse~2
respectively.  Spectra are fitted by
the CXE model with a solar wind velocity of 450 km sec$^{-1}$.  All
spectra show relatively strong emission from \OVII\ and the innermost
regions show the strongest \CVI\ and \CV\ emission.
\label{fig:specpanda}}
\end{figure}

\begin{figure}
    \centering
        \includegraphics[width=\textwidth]{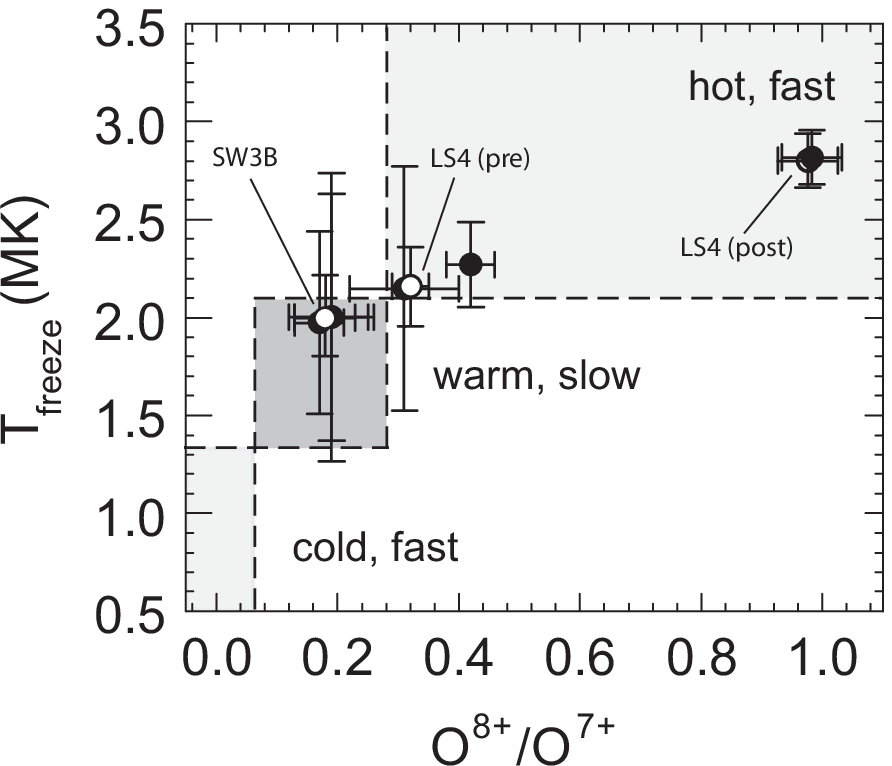}
    \caption{Spectrum derived ionic Oxygen ratios and corresponding
    freezing in temperatures from Mazzotta \e (1998). Dots with error
    bars indicate all comets observed with $Chandra$. 73P/B and LS4
    (pre- and post- breakup)
    are identified explicitly. LS4 post-breakup is nearly coincident
    with 153P/Ikeya-Zhang. The shaded
    area indicates the typical range of the slow solar wind ({\it Figure adapted from Bodewits \e 2007}).}
\label{fig:swox}
\end{figure}

\skipthis{
\begin{figure}
 \plotfiddle{fig11.eps}{-1.0in}{90}{360.}{500.}{-50.}{500.}
\caption{Comparison of two fits adding low excitation lines of Si to
  our nominal ``bigbox'' fit.  In the top figure, the inclusion of the
  Si lines lowers the \CV\ flux by 25\% while the $\chi^2$ of the fit
  remains $\sim$ 1.2 per d.o.f.  The lower figure shows the same data
  with the model flux from \CV\ held to be consistent with the solar
  wind abundance  (10\% of its value in the nominal fit).  
The  $\chi^2$ of this fit is insignificantly higher than the top model.}
\label{EXP}
\end{figure}
}

\begin{figure}
 \plotfiddle{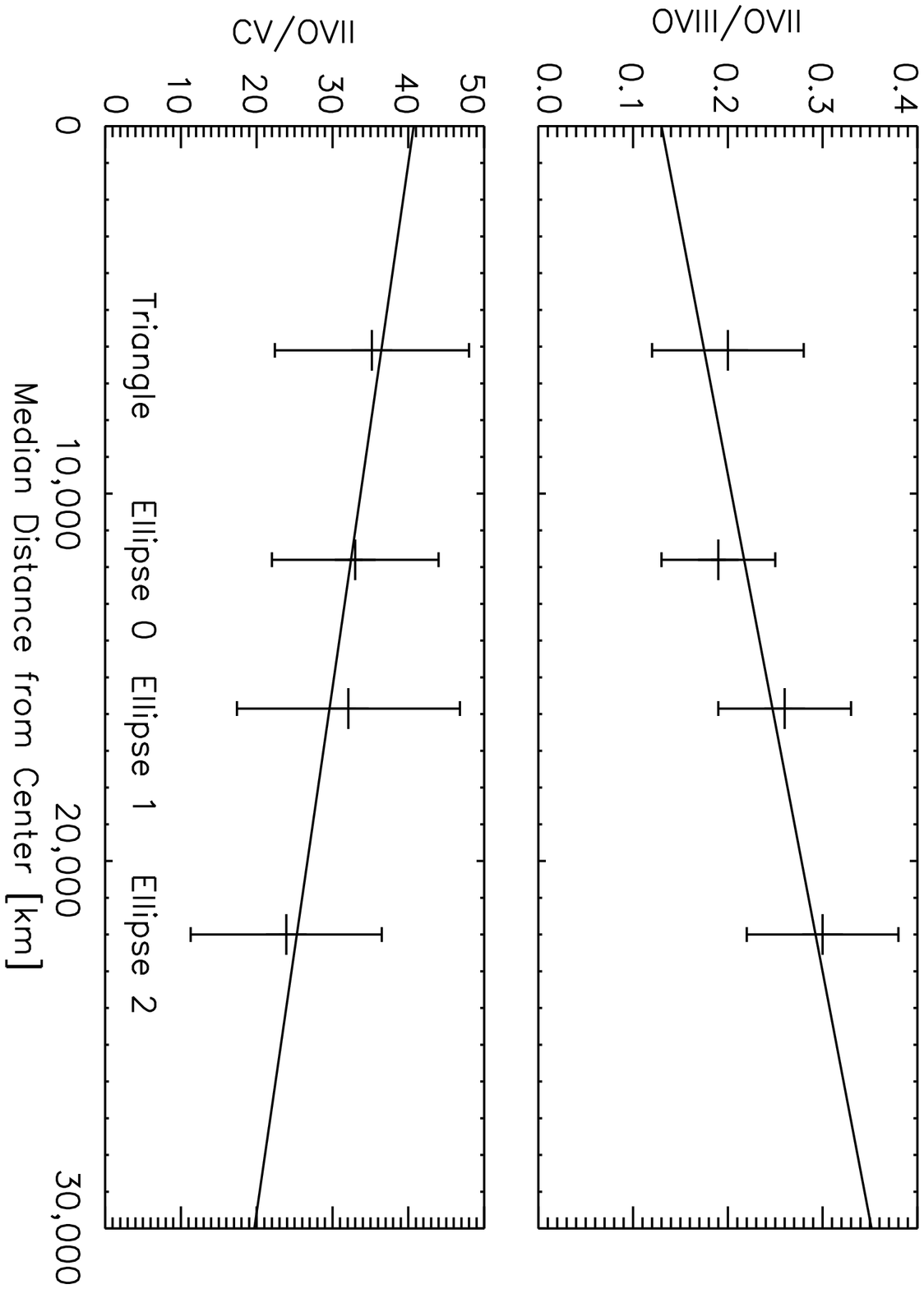}{-1.0in}{90}{360.}{500.}{00.}{300.}
\caption{Trends in the ionization ratios measured in the emission lines vs. distance from the comet's nucleus. 
  The top figure shows \OVIII/\OVII, the bottom \CV/\OVII.  Formal errors are included in the ratios.
  The X-axis shows the median distance of the measured points to the center of the ``Triangle" region.
  The straight lines are fitted to the data weighted by the errors in the measured ratios.
   \OVIII/\OVII\ is seen to increase with increasing distance while  \CV/\OVII\ decreases.
  }
\label{trends}
\end{figure}

\begin{figure}
    \centering
        %\plotfiddle{fig12.eps}{-1.0in}{90}{360.}{500.}{-50.}{500.}
         \plotone{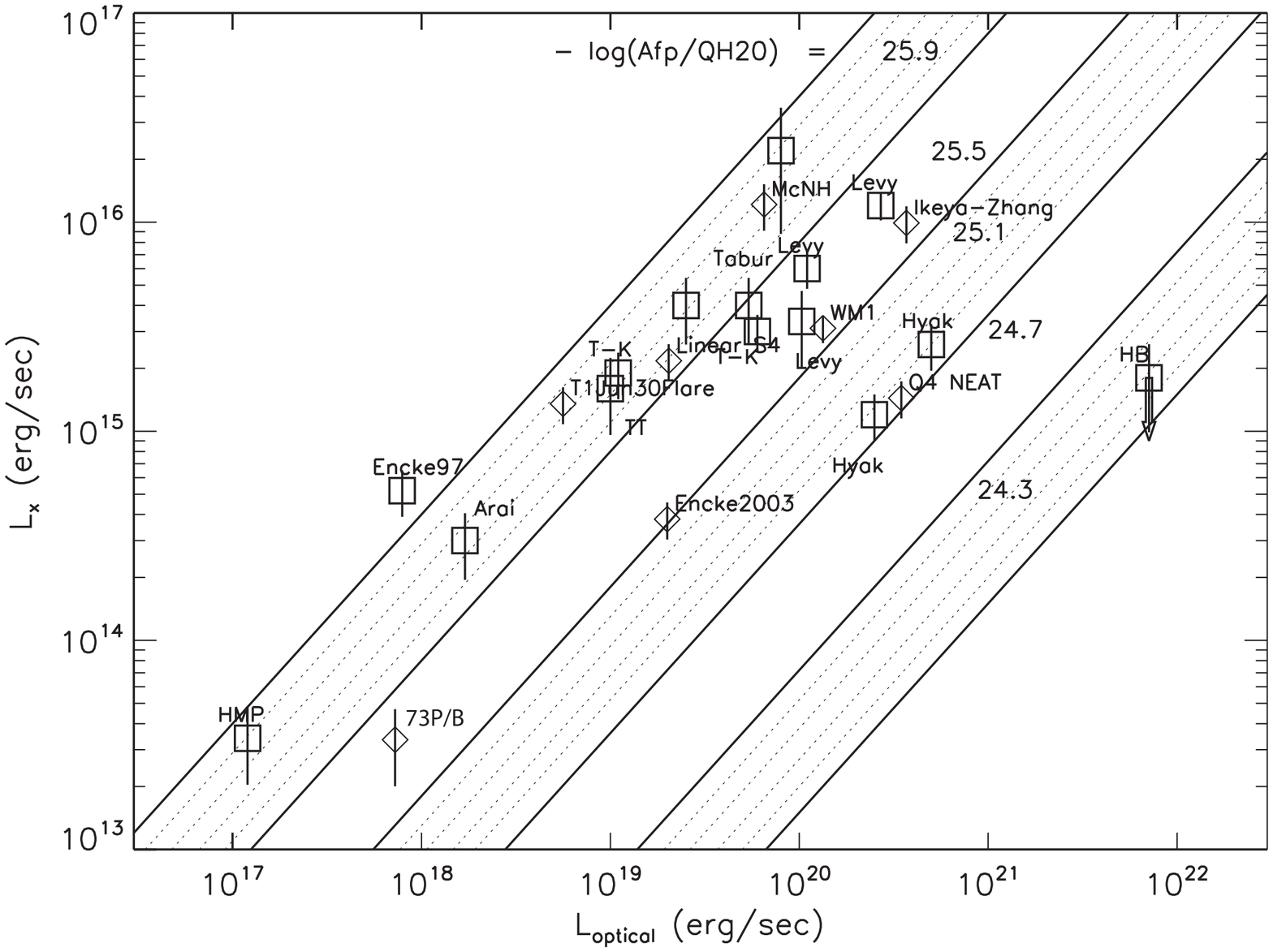}
%{Screenshot-1.eps}
    \caption{Plot of the X-ray versus optical luminosity for about 15 comets.  The diagonal lines represent constant dust to gas
    ratios for comet observed with $ROSAT$ until 1997. $Chandra$
    observed comets are shown with diamonds, other X-ray observed
    comets are shown with squares.  The typical uncertainty of 0.2 mag (2 sigma)
     in the optical flux translates to an uncertainty of $\sim$ 20\% in L$_{optical}$. 
     At lower X-ray luminosities comets seem to lie along a
    constant  dust to gas ratio. X-ray luminosities appear to reach a
    maximum around $L_x \approx 10^{16} \erg\ \ps$.   The unexplained faintness of comets (Hyakutake, NEAT, Hale-Bopp) at very high D/G ratios may argue for a mechanism of X--ray attenuation, like CXE between solar wind ions and dust particles, that favors Auger electron emission instead of X--ray emission, 
\label{LXLOPT}}
\end{figure}

\begin{figure}
    \centering
        \plotone{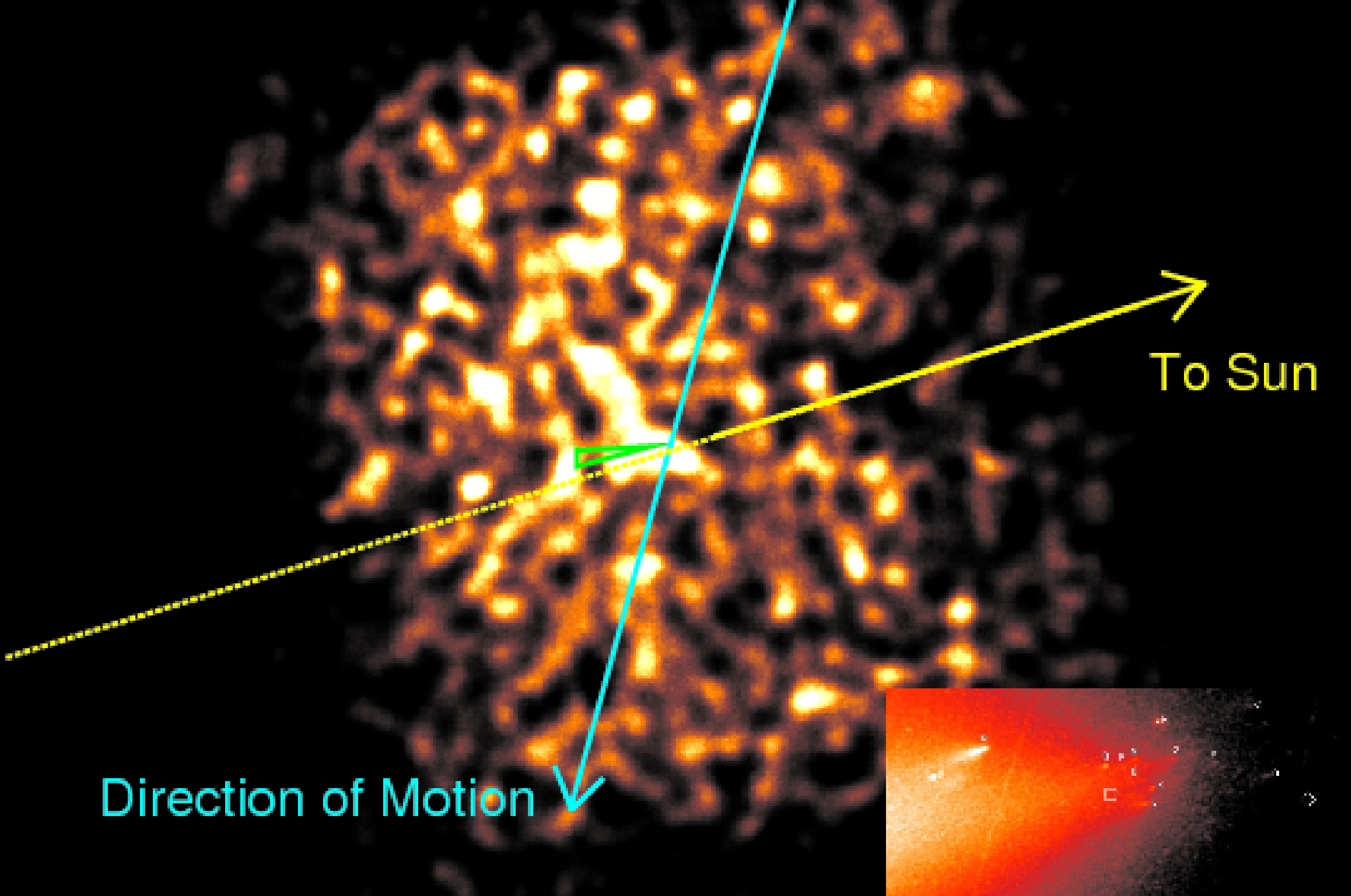}
    \caption{Smoothed image of CCD S3 from the observation of LS4 on 1 August 2000. In this image north is up and east is to the left.
     Counts from energies between 0.3-1.2 keV are shown. Squared scaling is used.
     Data are binned  and then smoothed using a Gaussian kernel, the final effective pixel size is $\sim
     13$\arcsec.  The Sun line (yellow) and direction of
     motion (cyan) are
     indicated, as is  the location of a small grouping of large fragments (D $>$ 70 m).  identified by
     Weaver \e (2001; green triangle). The nominal nuclear position is
     the intersection of the vectors.  Emission is brightest near the nominal
     location of the nucleus and in the (northeast) quadrant
     trailing the comet and away from the Sun. {\bf Inset:} Figure~1 (middle) from Weaver \e 2001. An HST/WFC image of LS4 taken 5 August 2000.  The diamond gives the predicted position of the original nucleus using the JPL-87 orbit solution, and the square shows the predicted position of the nucleus using the JPL-95 orbit solution. Their separation is 19.3\asec. 
\label{img:ls4}}
\end{figure}

\begin{figure}
    \centering
         \plotfiddle{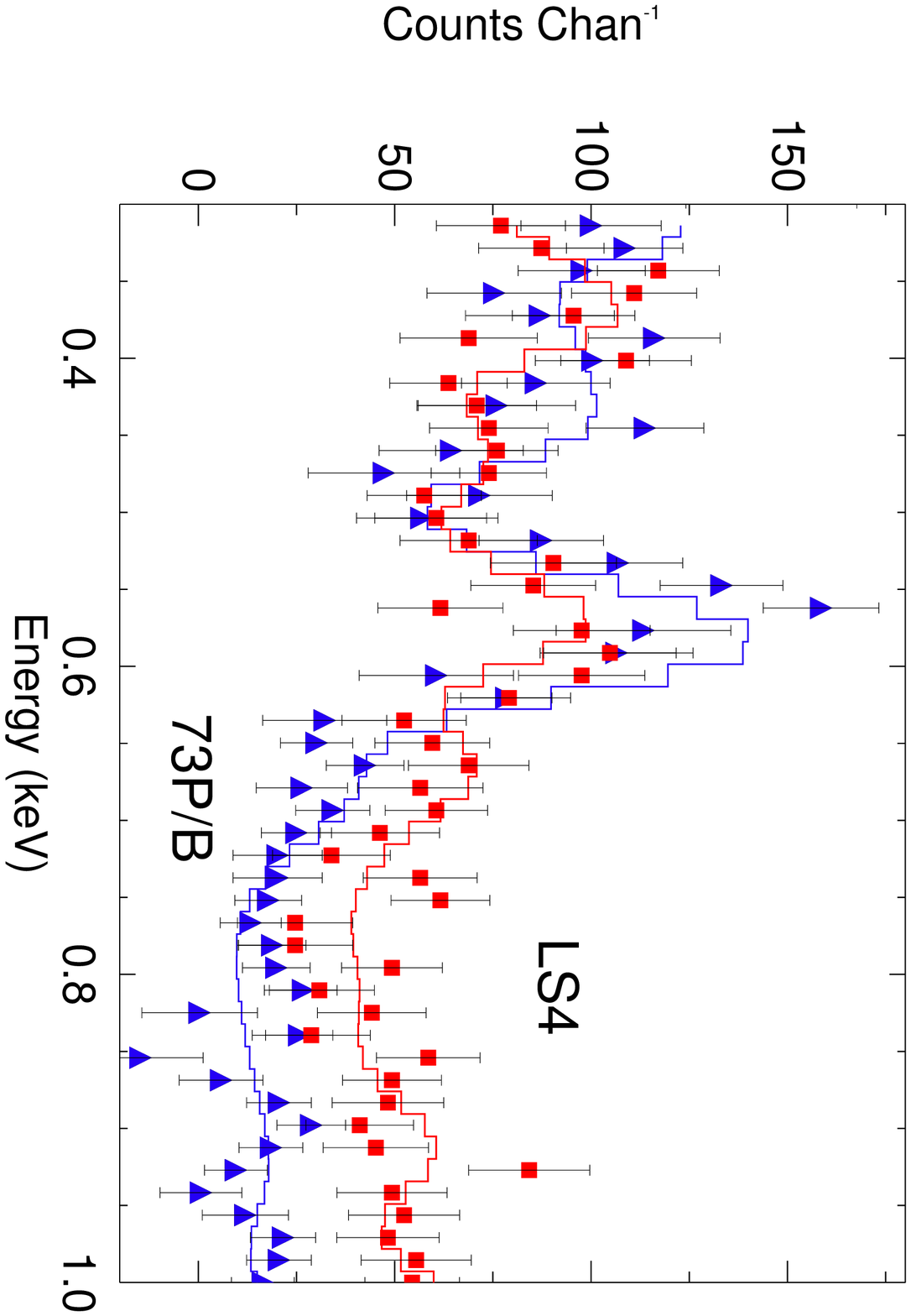}{0.0in}{90}{360.}{500.}{-50.}{-500.}
    \caption{The spectra of LS4 (Red squares) and 73P/B (Blue and triangles)
    from 0.3-1.0 keV. LS4 shows stronger \CVI, \OVIII\ and \NeIX\
    (0.37, 0.65 and 1.02 keV respectively) than
    73P/B despite the fact that it had ceased to exist as a unique body
    over a week earlier.
\label{fig:LS4spec}}
\end{figure}

%%%%%%%%%

\end{document}